\documentclass[11pt,a4paper]{article}

\usepackage[T1]{fontenc}
\usepackage[utf8]{inputenc}
\usepackage[british]{babel}
\usepackage{lmodern}
\usepackage{microtype}
\usepackage[margin=2.5cm]{geometry}
\usepackage{graphicx}
\usepackage{longtable,booktabs,array}
\usepackage{caption}
\usepackage{newunicodechar}
\usepackage{xurl}
\usepackage[hidelinks]{hyperref}
\usepackage{orcidlink}

\newunicodechar{ṁ}{\.m}
\newunicodechar{·}{\kern-0.15em\textperiodcentered\kern-0.15em}
\newunicodechar{→}{\ensuremath{\rightarrow}}

\newcommand{\nl}[1]{\raisebox{-\dp\strutbox}{#1}}
\newcommand{\nlheight}{\dimexpr\ht\strutbox+\dp\strutbox\relax}

\title{One Future Fits All: Text-to-image AI and the Narrowing of Global Imagination}

\author{\parbox{\textwidth}{\centering\normalsize
\mbox{Stephanie Diepeveen\textsuperscript{1}\,\orcidlink{0000-0002-5468-9312}},
\mbox{Somnath Batabyal\textsuperscript{2,*}\,\orcidlink{0000-0002-0382-8351}},
\mbox{Edward Ademolu\textsuperscript{1}\,\orcidlink{0000-0002-5122-6232}},
\mbox{Peter Chonka\textsuperscript{1}\,\orcidlink{0000-0001-9871-1377}},
\mbox{Gabriele de Seta\textsuperscript{3}\,\orcidlink{0000-0003-0497-2811}},
\mbox{Iginio Gagliardone\textsuperscript{4}\,\orcidlink{0000-0002-2878-7963}},
\mbox{Elisabeth Gunawan\textsuperscript{5}\,\orcidlink{0009-0009-8489-7866}},
\mbox{Fabio Gygi\textsuperscript{2}},
\mbox{Yidnekachew Haile\textsuperscript{6,7}\,\orcidlink{0000-0002-7927-7357}},
\mbox{Yenn Lee\textsuperscript{2}\,\orcidlink{0000-0002-0708-2948}},
\mbox{Chipo Mapondera\,\orcidlink{0009-0004-5078-9254}},
\mbox{Craig Ryder\textsuperscript{8}\,\orcidlink{0000-0003-4624-9293}},
\mbox{Matti Pohjonen\textsuperscript{9}\,\orcidlink{0000-0003-1702-6402}},
\mbox{Michiel Baas\textsuperscript{10}\,\orcidlink{0000-0003-4405-146X}}\\[1.2em]
{\small\begin{tabular}{@{}c@{}}
\textsuperscript{1}King's College London, UK\\
\textsuperscript{2}SOAS University of London, UK\\
\textsuperscript{3}University of Bergen, Norway\\
\textsuperscript{4}University of the Witwatersrand, South Africa\\
\textsuperscript{5}School of Music, English and Theatre, Goldsmiths, University of London, UK\\
\textsuperscript{6}Teesside University, UK\\
\textsuperscript{7}Royal Holloway, University of London, UK\\
\textsuperscript{8}LMU University of Munich, Germany\\
\textsuperscript{9}University of Helsinki, Finland\\
\textsuperscript{10}Leiden University, The Netherlands\\[0.6em]
\textsuperscript{*}Corresponding author: \href{mailto:sb127@soas.ac.uk}{sb127@soas.ac.uk}
\end{tabular}}}}

\date{}

\begin{document}

\maketitle

\begin{abstract}
Persistence of algorithmic bias is well established in scholarship, reflecting, reproducing and reshaping inequalities and injustices. Yet, as technology changes, we cannot assume that inequalities will necessarily be reproduced in the same or similar forms. This article explores the reproduction of bias in text-to-image generative AI, looking at the tool, Microsoft Bing Create, based on DALL·E 3. It draws on the results of a collaborative exploratory study of 907 image-based outputs of text-to-image generation in 15 languages and concerning 12 countries. It interrogates the gaze of generative AI models, and how scholars can indirectly study their opaque operations. To this end, we utilise the notion of `synthetic probes' (de Seta, 2024) to iteratively explore correlations of inputs and outputs within the context of place, language and power. We find striking simplicity in the scope of possibilities through which Bing Create considers the current and future city. Stripping out (explicit) indications of conflict, poverty and politics, we are left with images whose localised diversity harkens back to versions of Orientalist imaginaries. Images of the future emerge as one civilising mission or modernisation narrative, with all places headed to one future of hovercrafts, urban gardens and solar panels, encased in a composition of shades of blue and green. We argue that the visual outputs of this generative AI tool reflect and reproduce a simple but pervasive and persistent colonial/developmental worldview. We explore explanations for this emergent narrative and its implications, suggesting the threat of generative AI is not in its complexity or intelligence, but rather its banality. With this, we propose banality not as mere repetition, but the systematic narrowing of representational possibility: a convergence on tourist-friendly, depoliticised, development-saturated visual grammars that smooth contradiction and foreclose plural futurisms.

\end{abstract}

\section{Introduction}

In 2025, a series of incidents with Elon Musk's Grok Large Language Model (LLM) brought to the fore concerns over the extent to which human intervention can direct the scope of a model's outputs. In May, Grok began generating outputs referring to conspiracies about white genocide in South Africa in response to unrelated prompts (Kerr, 2025). In July, Grok was recorded providing antisemitic outputs, even referring to itself as `MechaHitler'.\footnote{For discussion of Grok's generation of antisemitic outputs, see Wong (2025).} Elon Musk, founder of X.ai the company behind the Grok LLM, has publicly stated his intention for Grok to be open to `politically incorrect' viewpoints, indicating a willingness to programme rules into the model to delineate outputs in line with his normative outlook (Wong, 2025).

These incidents sparked debates over how artificial intelligence (AI) alignment methods -- which intervene in the model's learning aiming to restrict what outputs are generated -- can and should be used (Foulds et al., 2025). On one hand, they reaffirmed the unknowability-- and at least what appeared to be some unpredictability -- in how the model relates input prompts to outputs, visible in unsolicited outputs. On the other hand, updates to Grok linked to Elon Musk's values indicate a growing spectre of opportunities for humans to introduce selectivity into models.

While Grok is a text system, the controversy surfaces a wider issue that structures text-to-image tools: how alignment, safety constraints and commercial design choices shape what a model can plausibly depict and, crucially, what it consistently cannot. As an interdisciplinary group of researchers with a shared interest in technology and society in the majority world\footnote{In invoking the terms majority world and Global South (versus Global North), we recognise the challenges of adopting such terms given diversity and dynamism within and across countries. We use these terms to indicate our orientation to studying AI from places and people that have tended to be marginalised or decentred in scholarship.}, we propose that these debates over the potential for humans to intervene in how models generate outputs, while technologically important, are problematic in their narrowness. They neglect fundamental constraints tied to the knowledge hierarchies in which generative AI models -- and their data, design and engineers -- sit.

This position is founded on a collaborative experiment undertaken by fifteen researchers, interrogating the outputs of a text-to-image based generative AI tool (Microsoft Bing Create, based on DALL·E 3) in 2024 and 2025. This article is structured around two questions: How do text-to-image systems reflect and/or (re)produce colonial/developmental imaginaries in representations of place? How do such outputs help us conceptualise the nature, operations and possibilities of colonial power in contemporary generative AI infrastructures?

Inspired by ``synthetic ethnography'' (de Seta, Pohjonen and Knuutila, 2024), we explored what could be revealed through a systematic examination of the relationship between prompts and outputs. We contribute to scholarship on digital colonialism and AI bias by invoking Edward Said's approach to power and its representation. This entailed examining prompts and outputs from a Foucauldian view of power and discourse, in which representations of the world are socially organised and constitute ``imaginative geographies'' with cultural, social and material dimensions (Edward Said, in Urry and Larsen, 2011: 186). Analytically, we focused on the role of discourse in image generation. This directed our inquiry to the archaeology of datasets and the ``mode of discourse with supporting institutions, vocabulary, scholarship, imagery, doctrines'' (Said, 1978: 1-2) that make possible and normalise their production, obscuring alternatives.

We identify a disjuncture between leaps in computation with generative AI, and constraints posed by the discourse `encoded' into dominant models. An Orientalist lens reveals how such outputs repeatedly regurgitate `Orientalist' representations of places. We show that under our prompt protocol across multilingual cases, Bing Create/DALL·E 3 privileges Orientalist and developmental visual grammars, converging on a narrow repertoire of representations and futures. This is reflected in an inability to capture the plurality of places and experiences, and in representations of the future that depict aspirations of an Orientalist imaginary, obscuring contradictions.

We make three broad claims. First, the datasets used to build the AI tool are biased, shaped through Orientalist worldviews. Second, there has been little chance for alternatives, as dominant knowledge systems are embedded in power, racism and marginalisation. Third, the consequence of the above two: when the AI tool is prompted to generate images of our collective future, that prediction is not only banal and stereotypical, it is also logically impossible and dangerous. If taken seriously, it would further the harm already caused by colonial/developmental projects. Our concern with generative AI therefore lies not in existential concerns over autonomous computing systems. Rather, it is in the amplification of injustices that restricts the world to a narrow set of representations and masks their contradictions and destructiveness.

\section{Critical and decolonial studies of AI}

Decolonial and critical scholarship has already brought a sceptical lens to what Natale (2021) has called `magical thinking' about AI, in which AI is surrounded by a myth of being more than `artificial intelligence', harkening back to Elish and boyd's (2018: 35) identification of a rising ``technological myth centred around the possibility of creating thinking machines.'' Our study builds on decolonial scholarship exploring historically-rooted inequalities and biases in AI (Adams, 2021; Adams et al., 2023; Gorska and Jemielniak, 2023; Lewis et al., 2024; Mohamed et al., 2020).

The latest in AI technology sits within a history of colonialisation and its continued presence in dominant political, economic and knowledge systems (Mohamed et al., 2020; Couldry and Mejias, 2019). This is evident in the production of generative AI (Muldoon and Wu, 2023), from inequality in financing, decision-making and algorithmic design (Mohamed, Png and Isaac, 2020), to configurations of labour and value extraction (Anwar and Graham, 2020). Notions of intelligence within which AI systems are seen to excel are contingent upon colonial knowledge hierarchies (Adams, 2021, Lewis et al., 2024), visible in the fear of the potential for a more intelligent `thing' to colonise a less intelligent being/society (Cave, 2020).

Data are digital traces of human interactions, with their biases visible in what is represented and what isn't (Benjamin, 2019; Chonka et al., 2023). Biased data are reflected in biased algorithmic outputs (Buolamwini and Gebru, 2018; Noble, 2018). More complex designs and greater compute do not erase such biases in machine learning. Studies conducted on LLMs extend the empirical evidence of data-related biases (Gorska and Jemielniak, 2023; Locke and Hodgdon, 2024). Large scale audits of generative AI models consistently find multiple kinds of biases, which result in a specific ``silicon gaze'' shaped by the ``positionalities and power asymmetries of its training data, designers, and platform owners'' and perpetuate centuries of global inequalities. (Kerche et al., 2026: 1).

Syed Mustafa Ali argues ``it might be more useful to think about AI \emph{as} colonial in and of itself, since this raises the vexing question as to whether AI can be decolonised'' (2024: 648). Ali's proposition is a useful starting point for inquiry. AI-generated imagery relies on pre-existing visual and cultural data (Alenichev et al., 2023). As AI systems reproduce images (and words), injustices and racism of the past are regurgitated. Efforts to improve training datasets to mitigate such bias may be undermined by the power of dominant global models, and their recursive impact on digital publics. With the near unfathomable (and increasing) computing powers behind AI systems, these reproductions are multiplying at exponential rates, generating more AI content -- or AI `slop', which feed into future outputs, entrenching historical power imbalances at this crucial moment of generative AI's development and popularisation. Whatever algorithmic gymnastics, and however probability models are reshaped, these models carry past injustices.

We also acknowledge AI is invested with possibility, from existential risk to opportunity. Equating AI with colonial pasts can bring a sense of determinism to its possibilities. Narratives surrounding AI point to the unknowability of its most accurate predictive outputs. There is a sense of an ``unstoppable computerisation of advanced industrial societies'' (Carabantes, 2020:309). This direction of travel appears, at least at the time of writing, to be underpinned by a paradox whereby the AI models with the best predictive accuracy are the most opaque (Carabantes, 2020).

We are interested in interrogating generative AI image outputs given the technology's colonial foundations. Are generative AI tools bound by the limitations visible in less complex machine learning models? Or might the hype around AI tools' autonomy and unknowability indicate space for alternative -- even new -- imaginaries of the majority world?

We approach our study from the premise established in decolonial studies that AI sits within colonial systems of knowledge (Ali, 2024; Couldry and Mejias, 2019). Equally, we create space to interrogate what might be new with generative AI models. As Catriona Gray (2023) argues in relation to data and colonialism, technological forms of exploitation layer upon historical colonialism, as opposed to being analogous.

\section{Orientalism and AI}

Edward Said offers a path for inquiry into connections between representations of the world and the western-dominated systems of power in which they sit. In utilising Said's analytical perspective, we complement Kotliar (2020) who has invoked Orientalism to interrogate power in profiling algorithms, identifying the deep rootedness of Orientalist discourse, while also conceptualising algorithmic design as an intercultural process with power flowing in multiple directions. Our inquiry expands on this, considering systems of knowledge and the operations and outputs of a text-to-image generative AI tool.

In \emph{Orientalism,} Said explores how representations of people and places dating to the mid-18\textsuperscript{th} century have come to define systems of power. He explains how the `west' has defined itself in relation to the rest of the world through a series of binaries, bound up in colonial encounters. Orientalist representations are characterised by the agency of those who produce them: the European (later American or western person). This results in representations of places and people constructed ``through and by virtue of the European imagination'' (Said, 1978: 56) that are founded on an imagined -- and constructed - divide between the Occident/West and the Orient/East `other'.

We use `Orientalism' less as a geographic label than as a representational logic: a way of producing places as consumable, legible \emph{elsewheres} through a narrow repertoire of tropes, even when the place in question sits within the global North. Orientalist imaginaries persist through representational infrastructures, including visual economies of development and humanitarianism (Chouliaraki, 2006; Dogra, 2012; Lidchi, 1999); global film and popular culture (Hall, 1992; Keenaghan and Reilly, 2017; Lewis et al., 2014); the tourist industry (Arjana, 2025; Kothari, 2015); and social media (e.g. Ademolu 2024; Chonka, 2019; Richey, 2016; Schwarz and Richey, 2019), each of which has long generated images of places in the global majority and organised how they become legible to distant audiences.

In looking to manifestations of Orientalist imaginaries, it is useful to place Orientalism in conversation with scholarship on humanitarian, development and ethical visuality, which examines how regions and communities are rendered as objects of distant spectatorship through depoliticised and consumable visual grammars. Work on the visual politics of `distant suffering' has shown how images organise moral distance and political legibility, neutralising political history through readable scenes of need, resilience and redemption (Ademolu, 2023, 2025; Campbell, 2012; Chonka, 2019; Chouliaraki, 2006, 2010; Sontag, 2003). Read through this lens, the visual patterns we identify are not only `Orientalist' in Said's sense, but part of a larger visual economy in which the global majority is routinely made visible through sentimental smoothing, depoliticised inclusion, and developmental promise (Dogra, 2012; Lewis et al., 2014).

Two versions of these visual patterns are important to our study given the images generated. First, John Urry and Jonas Larsen show how mass tourism, emerging from 1840s post-industrial Britain, produces versions of Said's ``imaginative geographies'' that depict places outside of one's (western) home within ``a seductive mix of reverie, reality and fiction -- of simultaneous `naturalisation' and `fictionalisation'\,'' (Urry and Larsen, 2011: 195-6). Second, projects of development remain underpinned by essentialist views of a `developed' world, and a less developed world that imitates, repeats and seeks to catch up, whether framed through civilising missions, Rostow's (1960) stages of growth, or recent narratives of digital development (Couldry and Mejias, 2019; Escobar, 1995; Ferguson, 1994; Sachs, 2019; Toyama, 2015). What becomes distinctive in the context of text-to-image generative AI models is not the invention of these regimes, but their automation, scaling and recombination as platform-mediated defaults.

Orientalism is informed by Foucault's theory of discourse\footnote{Said had an ambivalent and changing view of Foucault. Said's early work owes -- and recognises - the influence of Foucault's conceptualisation of discourse in explaining how Orientalist representations relate to global power structures, thereby extending Foucault's idea of discourse to global majority contexts (Racevskis, 2005).} and Said connects knowledge production to broader forms of western domination (Chatterjee, 2018; El Aidi and Yechouti, 2017). For Foucault, discourse is a representation that is produced by power and reproduces that power. With Orientalism, connections between knowledge and power explain the ubiquity and production of particular imaginaries of places and people globally.

This focus on discourse is picked up within wider postcolonial scholarship exploring visual patterns of Orientalism. Urry and Larsen (2011) tie the ubiquity of tourist images not only to the tourist, but also the actions and presentations of people who are the object of the tourist gaze. Escobar (1984) extends Foucault's discourse to understand the normalisation of development in the post-WWII era. He weaves together the roles of ideas, institutions and processes in maintaining development, from everyday subjections to grand strategies, to the industry's professionalisation, institutionalisation and adaptability to new problems.

Exposés and critiques of the discourse of development bring attention to a contradiction at the heart of representations of the west and the other. This is visible in the colonial project, whereby colonialism - especially in late-stage British and French colonialism in sub-Saharan Africa - promised civilisation while simultaneously denying it to indigenous communities by categorising them as subjects and exploiting resources for the metropole's profit (cf. Mamdani, 1996). This contradiction is also visible in the discourse of development: promising positive change while failing to contend with forms of exploitation and inequality that sustain global capitalism and advanced industrial economies (Escobar, 1984; Sachs, 2019). The world is divided into haves and have nots, with development as the promise to the have nots (Sachs, 2019). Yet, this promise is an impossibility, whether prevented by unequal terms of global trade or the unsustainability of advanced industrial economies (Sachs, 2019) - even without considering resource demands of advanced computing (Crawford, 2021).

Framing our analysis in Orientalism enables us to interrogate both the content of the images generated and how they connect to systems of power. Looking to generative AI, this expands our inquiry to the ``mode of discourse with supporting institutions, vocabulary, scholarship, imagery, doctrines'' (Said, 1978: 1-2) that normalise images' production and erase alternatives.

\section{Methodology}

Our study adopts a collaborative approach grounded in ``synthetic ethnography'' (de Seta, Pohjonen and Knuutila, 2024). Synthetic ethnography extends qualitative, participatory and dialogic ethnographic methods to sociotechnical systems, treating generative AI models as empirical sites through which broader cultural, political and representational formations can be examined (Plantin et al., 2018; van Dijck, 2013). During a workshop in June 2024, we experimented with a range of AI systems, including LLM-based chatbots and text-to-image and text-to-video models, to explore how systems mediate imaginaries of places and futures.

Given our diverse linguistic, regional and disciplinary backgrounds, we converged on a comparative design to examine how text-to-image generative AI structures imaginaries across geopolitical contexts. Following a phase of exploratory prompting, we selected a single modality (text-to-image) of a widely accessible platform (DALL·E 3, accessed via Microsoft Bing Create). This design choice allowed us to treat Bing Create as a mass-market representational infrastructure rather than as a neutral model in isolation. Our prompting strategy drew on approaches such as ``ambiguous prompting'' (Colombo et al., 2023), which are intended to surface how generative systems resolve indeterminate and underspecified requests, thereby revealing the representational grammars they privilege.

Working with a sample of twelve countries, we prompted the model in multiple languages per country. After a two-week testing period, prompts were entered on a weekly basis for a further twelve weeks. Each query produced four images, of which we downloaded the first (top left) image for analysis. Cases were selected pragmatically based on (i) authors' regional and linguistic expertise, (ii) coverage across multiple world regions and geopolitical positions, and (iii) the desire to include global north and global majority contexts and test whether representational narrowing travels across these categories. The multilingual design was inspired by writer/journalist Sakariye Cismaan's observation regarding divergent. Google Image search results for Mogadishu/Muqdisho (the English/Somali spellings of the capital city of Somalia).\footnote{See: \url{https://x.com/sakariyedhegnas/status/1167028614451384325}} This demonstrated how language can operate as a filter shaping what forms of place become visible and legible. By systematically varying language while holding prompts constant, our method treats language as an analytic lever for surfacing platform-mediated imaginaries.

We formalised three additive probes --- ``{[}country name{]}'', ``city'', and ``future'' --- reflecting a shared interest in urban contexts and futuristic imaginaries. These probes were translated into a selection of representative languages (Table~\ref{tab:data}) and applied through a standardised data collection protocol designed to minimise contextual carry-over effects (e.g., fresh chat windows, no customised system instructions). Translations were produced by researchers whose work focused on the language(s) and country.\footnote{In all but three cases (Sri Lanka, South Africa and Kenya) translations were done by researchers fluent in the languages. In these three instances, individuals {[}from ANONYMOUS, ANONYMOUS and ANONYMOUS{]}, fluent in each language, helped to complete and check the translations.} Where multiple plausible translations existed (e.g., `future' as temporal horizon versus imagined futurity), we agreed a selection of operational terms per language. In a few cases, researchers complemented the data protocol with additional prompts to explore patterns. This included prompting from different geographical locations (e.g. within and outside of a country), for North and South Korea as well as Korea, and for specific cities. Additional prompts were discussed collectively and used to refine our argument.

Analysis proceeded in two stages. First, country-level descriptive coding was undertaken by the researcher responsible for that country, documenting recurring motifs, compositional tropes, dominant signifiers and notable absences within the images. Second, findings were collectively interrogated in a workshop in November 2024, where comparison enabled the identification of shared visual grammars across contexts. The resulting analysis treats generative image outputs as discursive visual formations, allowing for examination of how platform-mediated AI systems systematically narrow and structure representations of place and futurity.\footnote{The database of images can be accessed at \url{https://github.com/objetpetitm/SOAS_AI_images}.}

\begingroup
\small
\setlength{\tabcolsep}{4pt}
\renewcommand{\arraystretch}{1.15}
\begin{longtable}{@{}>{\raggedright\arraybackslash}p{0.11\textwidth}
>{\raggedright\arraybackslash}p{0.15\textwidth}
>{\raggedright\arraybackslash}p{0.14\textwidth}
>{\raggedright\arraybackslash}p{0.10\textwidth}
>{\raggedright\arraybackslash}p{0.42\textwidth}@{}}
\caption{Data collection overview}\label{tab:data}\\
\toprule
\textbf{Country} & \textbf{Languages} & \textbf{Timeframe} & \textbf{Images}\footnote{The number of images varies by country due to a combination of factors: words that were common across languages; some missed weeks of translation; and researchers' decisions to extend the time period, generate outputs for more than two languages for a country, and/or to conduct additional tests.} & \textbf{Prompts used}\\
\midrule
\endfirsthead
\toprule
\textbf{Country} & \textbf{Languages} & \textbf{Timeframe} & \textbf{Images} & \textbf{Prompts used}\\
\midrule
\endhead
\bottomrule
\endlastfoot
Canada & French, English & 26 July -- 1 November 2024 & 75 & Canada; Canada city; Canada city future; Canada ville; Canada ville futur \\
China & Mandarin Chinese, English & 14 July -- 12 October 2024 & 84 & \nl{\includegraphics[height=\nlheight]{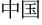}}; \nl{\includegraphics[height=\nlheight]{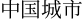}}; \nl{\includegraphics[height=\nlheight]{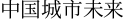}}; China; China city; China city future \\
Ethiopia & Amharic, English & 30 July -- 10 October 2024 & 36 & \nl{\includegraphics[height=\nlheight]{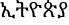}}; \nl{\includegraphics[height=\nlheight]{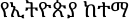}}; \nl{\includegraphics[height=\nlheight]{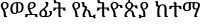}}; Ethiopia; Ethiopia city; Ethiopia city future \\
India & English, Hindi, Bengali, Assamese & 14 July -- 1 November 2024 & 78 & India, Bharat, Bharot, India city, bharot shohor, Bharat shaher, Bharat sohor \\
Indonesia & English, Indonesian & 16 July -- 11 November 2024 & 15 & Indonesia, Indonesia city, Indonesia City Future; Indonesia kota; Indonesia kota masa depan \\
Japan & Japanese, English & 26 July -- 1 November 2024 & 72 & \nl{\includegraphics[height=\nlheight]{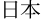}}; \nl{\includegraphics[height=\nlheight]{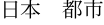}}; \nl{\includegraphics[height=\nlheight]{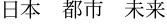}}; Japan; Japan city; Japan city future \\
Kenya & Swahili, English & 26 July -- 1 November 2024 & 90 & Kenya; Kenya city; Kenya city future; Kenya mji; Kenya mji wa kesho; Kenya mji wa siku zijazo\footnote{``Future'' was tested in two formulations in Kiswahili after discussion with Kiswahili linguistic scholars.} \\
Korea & Korean, English & 28 July -- 13 October 2024 & 92 & Korea [\nl{\includegraphics[height=\nlheight]{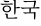}}]; Korea city [\nl{\includegraphics[height=\nlheight]{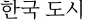}}]; Korea city future [\nl{\includegraphics[height=\nlheight]{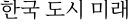}}]\newline
South Korea [\nl{\includegraphics[height=\nlheight]{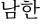}}]; Republic of Korea [\nl{\includegraphics[height=\nlheight]{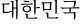}}]; North Korea [\nl{\includegraphics[height=\nlheight]{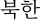}}]; People's Republic of Korea [\nl{\includegraphics[height=\nlheight]{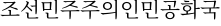}}] \\
Somalia & Somali, English & 10 July -- 31 October 2024 & 66 & Somalia; Somalia City; Somalia City Future; Soomaaliya; Soomaaliya Magaalo; Soomaaliya Magaalo Mustaqbal \\
South Africa & English, isiZulu, isiXhosa, Afrikaans\footnote{In South Africa, we chose to also prompt at the city level to explore potential variation when prompting specific cities. Additional languages (e.g.\ Afrikaans) were used.} & 26 July -- 1 November 2024 & 98 & South Africa; South Africa city; South Africa city future\newline
Iningizimu Afrika; Iningizimu Afrika idolobha; Iningizimu Afrika idolobha esizayo\newline
Cape Town; Cape Town future\newline
Stad Kaapstad; Stad Kaapstad toekoms\newline
Isixeko Sasekapa; Isixeko Sasekapa kwixesha elizayo;\newline
Johannesburg; Johannesburg future\newline
eGoli; eGoli esizayo \\
Sri Lanka & English, Sinhala, Tamil & 12 July -- 24 October 2024 & 99 & Sri Lanka; Sri Lanka city; Sri Lanka city future\newline
śrī laṁkāva; śrī laṁkā nagaraya; śrī laṁkā nagarayē anāgataya\newline
Ilaṅkai; Ilaṅkai nakaram; Ilaṅkai nakara etirkālam \\
Zimbabwe & English, Shona & 11 July -- 7 November 2024 & 102 & Zimbabwe; Zimbabwe dity; Zimbabwe city future\newline
Dzimbadzemabwe; Dzimbadzemabwe guta; Dzimbadzemabwe guta ramangwana \\
\end{longtable}
\endgroup

We implemented our study using Microsoft Bing Create\footnote{At the time of data collection, Bing Create claimed to support over 100 languages through Microsoft Translator. It also operated within AI standards and content safety approaches that sought to limit content production in relation to specific harms (e.g. violence, self-harm) and intended to promote fairness, including minimising stereotypes. Prompts infringing on these standards could be automatically blocked, with the user notified. None of our prompts were blocked. See: Microsoft (2022) ``Microsoft Responsible AI Standard, v2. General Requirements for External Release,'' June; and Microsoft (2024) \emph{Responsible AI Transparency Report 2024,''} May; and Microsoft. (2025). ``Harm Categories in Azure AI Content Safety,'' Microsoft, 21 November 2025. Available at: \url{https://learn.microsoft.com/en-us/azure/ai-services/content-safety/concepts/harm-categories?tabs=warning} (Accessed: 20 January 2026).}, a tool that generates images through DALL·E 3 from natural language descriptions. Our choice was informed by the popularity and availability of DALL·E at the time, and its relatively advanced text-to-image model. It was also accessible across researchers with varying technical experience, making it an effective base for collaborative manual data collection. Because Bing Create may expand prompts invisibly, our object of analysis is not `DALL·E in isolation' but the user-facing Bing Create assemblage (prompt → platform mediation → image output). This matters analytically, as it is precisely this layered mediation that shapes what kinds of place and futurity become depictable.

\section{Results}

Over three months, we produced 907 images. Reflecting our synthetic ethnographic approach, we invoke depictions from different sets of images to illustrate our findings. We first discuss images associated with the present-day, with the prompt structures {[}COUNTRY NAME{]} and ({[}COUNTRY NAME{]} + ``city''), and then when the prompt ``future'' was added, noting its strong effect on outputs. While there were some cases when the tool did not recognise the language prompt (e.g. in Shona, Assamese and Bengali), generally we find language had little consistent effect on image composition. Unless otherwise specified, findings discussed are common to local language and English prompts.

\subsection{Orientalist data equals Orientalist outputs}

\begin{figure}[htbp]
\centering
\includegraphics[width=0.55\textwidth]{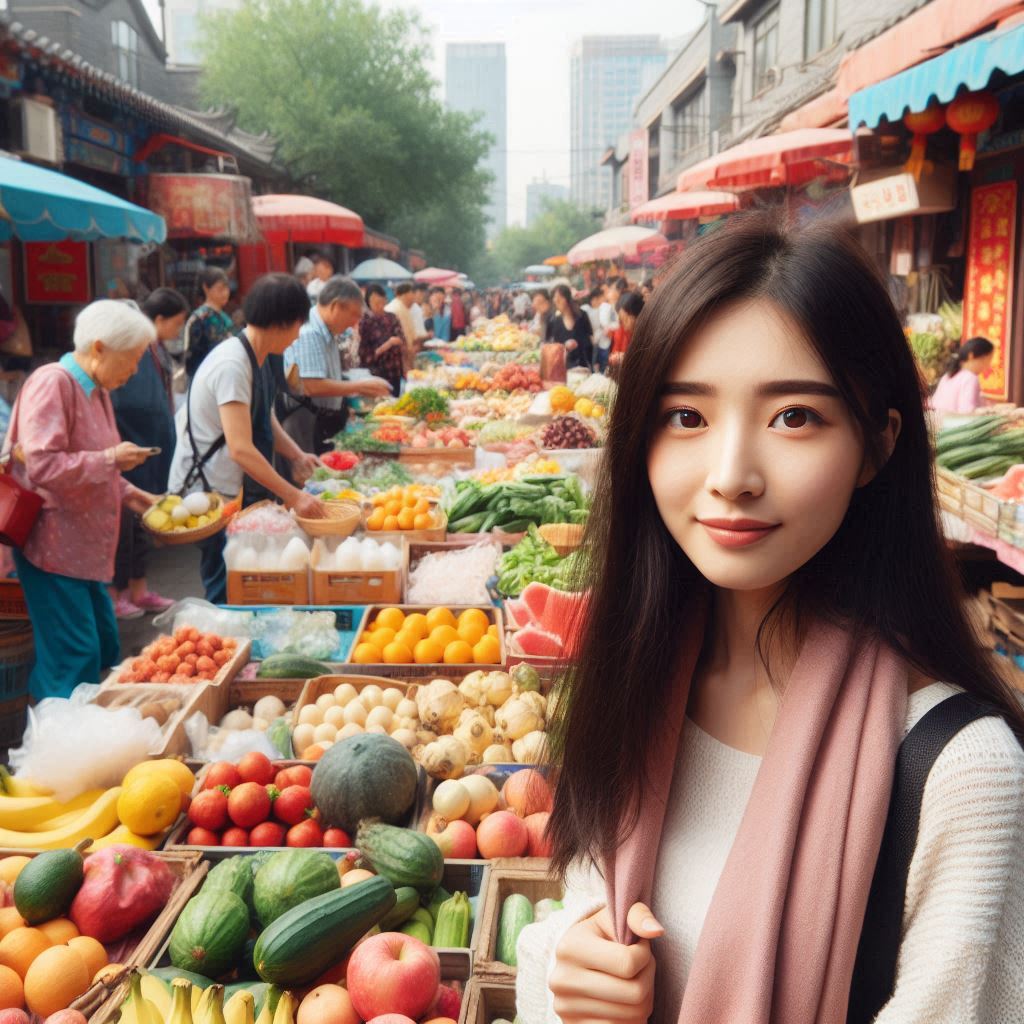}
\caption{``China + city'' (28 July 2024)}
\label{fig:1}
\end{figure}

\begin{figure}[htbp]
\centering
\includegraphics[width=0.55\textwidth]{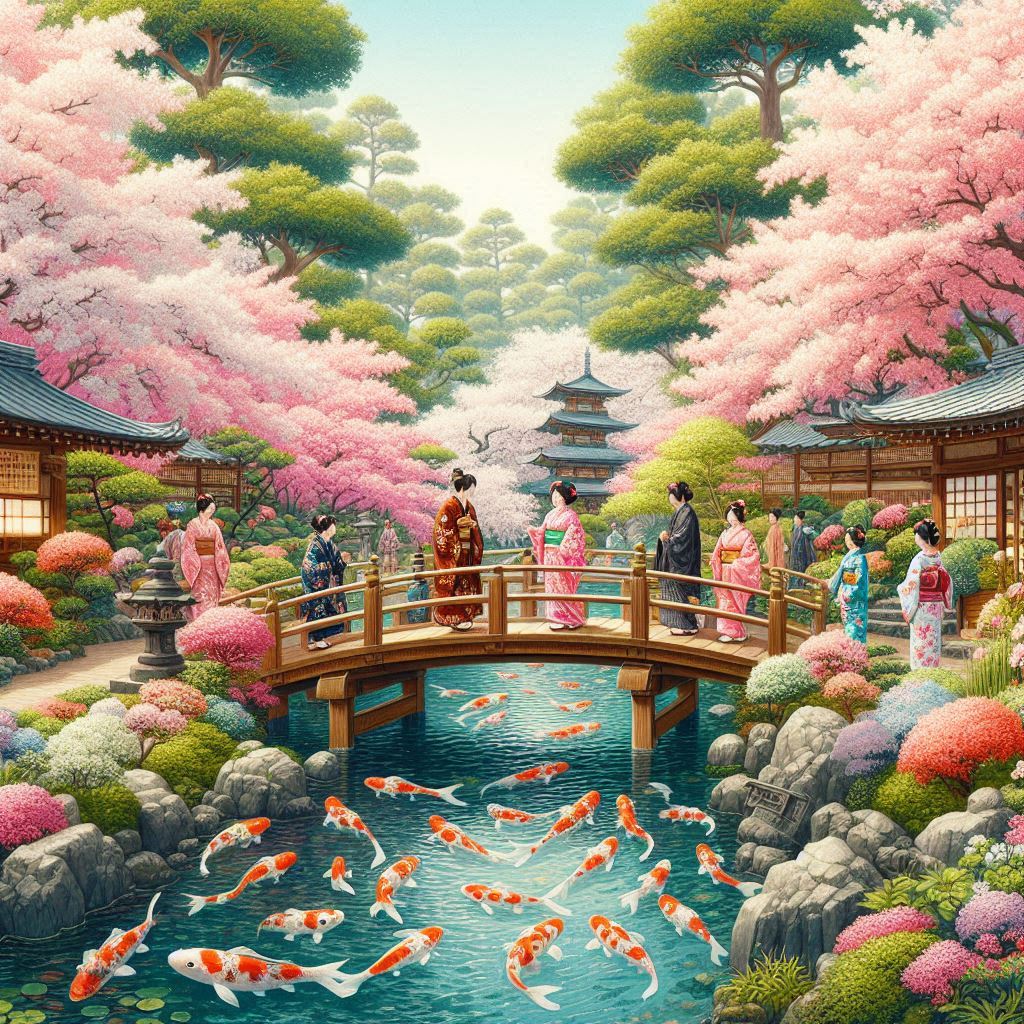}
\caption{``\nl{\includegraphics[height=\nlheight]{scripts/ja1.pdf}}'' {[}Japan{]} (16 September 2024)}
\label{fig:2}
\end{figure}

The most striking finding was the consistency of what the tool generated -- with a few signifiers for each country, such as lanterns and neon lights for China (Figure~\ref{fig:1}), or cherry blossoms and traditional kimono dress for Japan (Figure~\ref{fig:2}). Colonial histories and global hierarchies loom large, indicating that the tool may echo patterns of epistemic dominance. 

If Orientalism is a way of seeing and representing the world that is constructed by the West, looking across the images, we propose that the Bing Create tool constructs the world in Orientalist terms, not necessarily malevolently, but because the data that it relies on to construct probabilities, is Orientalist. Since the colonial era, much of the corpus of what is recorded as history has been observed and recorded by the West. The production of digital data has been disproportionately a product of western countries, where much of the world's computing capacity, connectivity networks, and resources for knowledge production lies. In probabilistic generative systems, training corpora sets the boundaries of the representational space: what is abundant becomes `probable', and what is absent becomes difficult to generate except as noise.

\begin{figure}[htbp]
\centering
\includegraphics[width=0.55\textwidth]{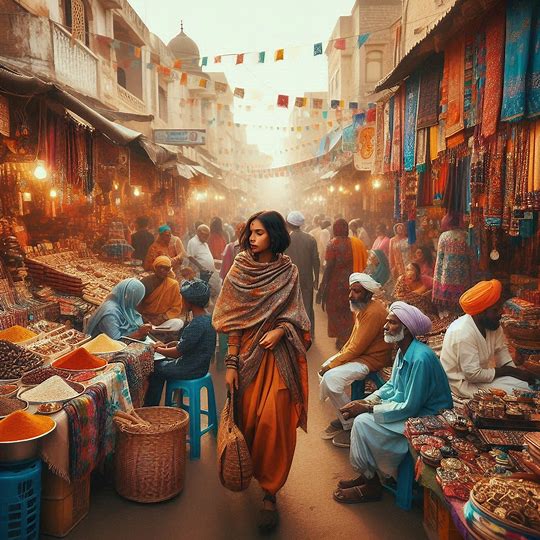}
\caption{``India'' (26 August 2024)}
\label{fig:3}
\end{figure}

\begin{figure}[htbp]
\centering
\includegraphics[width=0.55\textwidth]{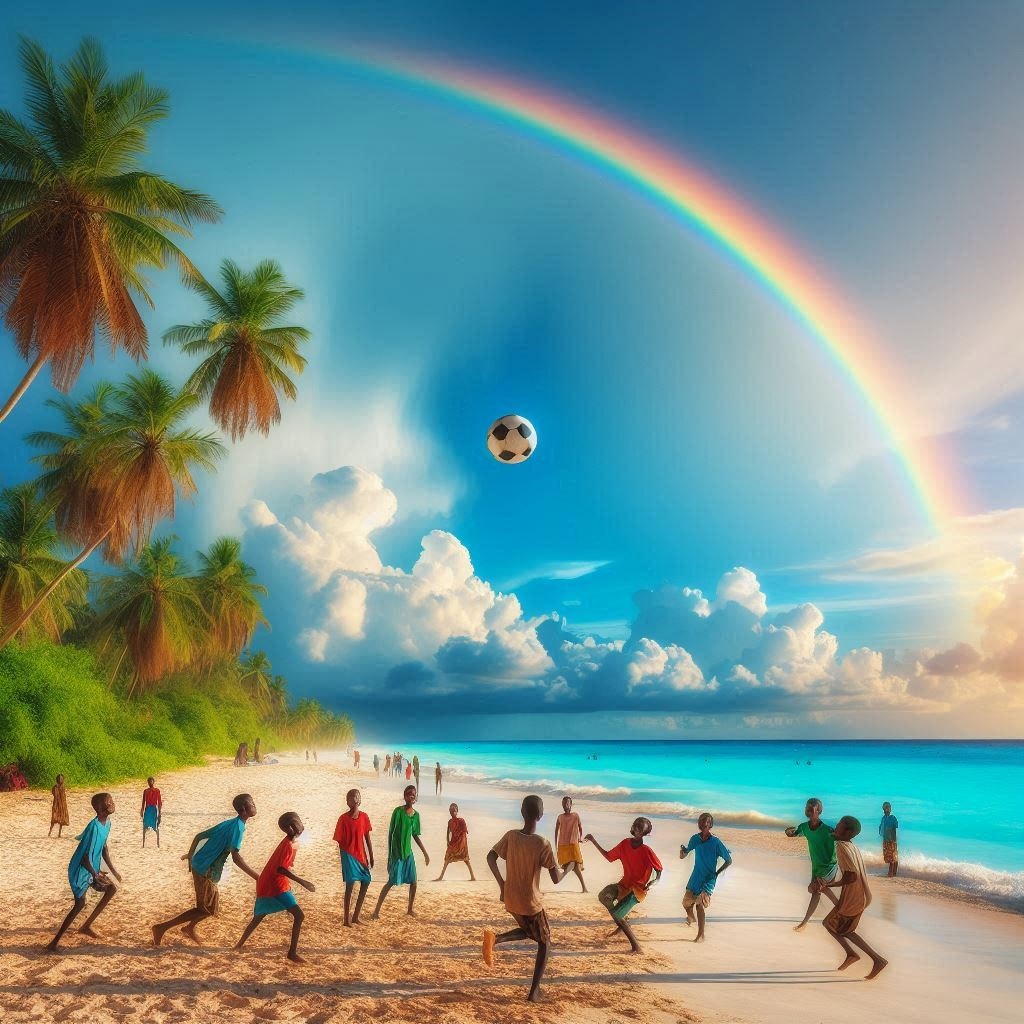}
\caption{``Soomaaliya'' {[}Somalia{]} (29 August 2024)}
\label{fig:4}
\end{figure}

\begin{figure}[htbp]
\centering
\includegraphics[width=0.55\textwidth]{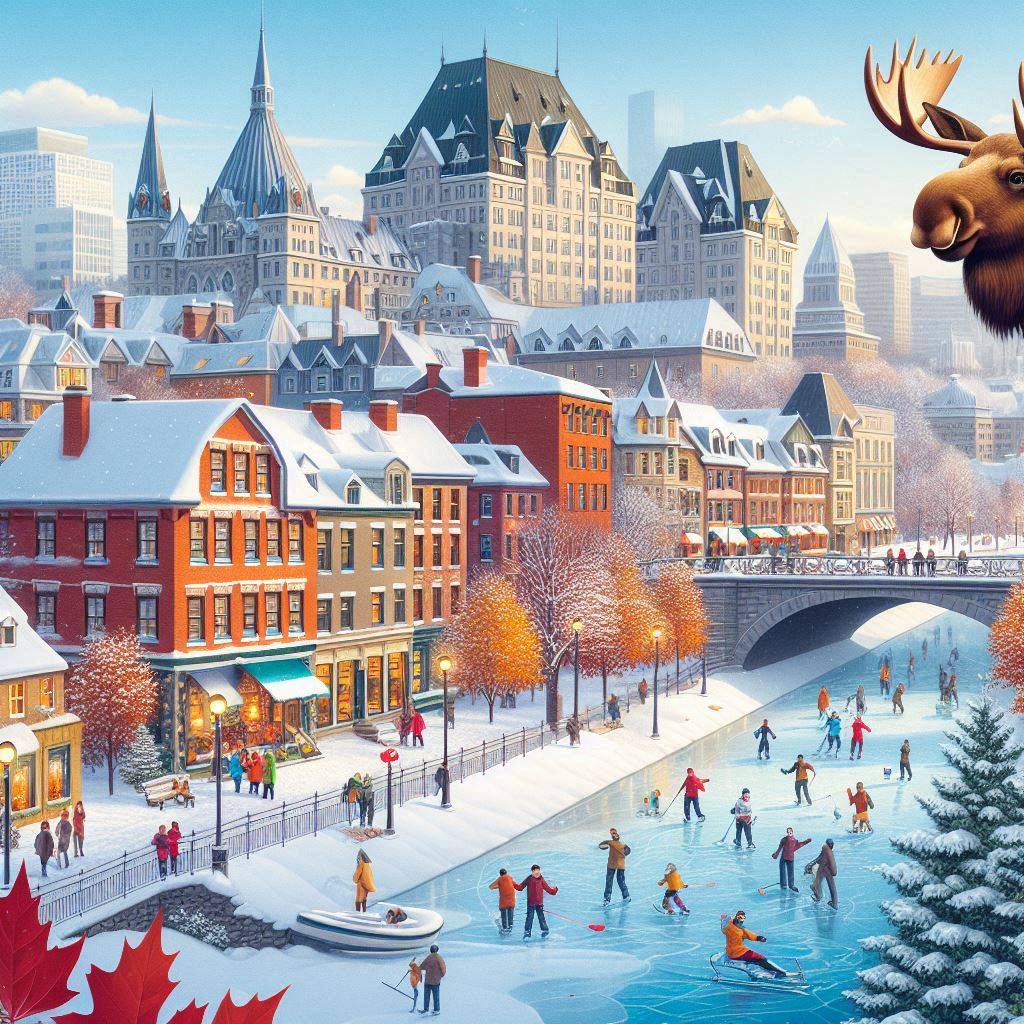}
\caption{``Canada + city'' (5 August 2024)}
\label{fig:5}
\end{figure}

Repeatedly, we find versions of Oriental images that exoticise and stereotype the majority world. Consider the images: a sari clad woman in a bazaar purporting to be India (Figure~\ref{fig:3}), or boys playing football on sandy beaches in Somalia, or was it South Africa (Figures~\ref{fig:4} and~\ref{fig:5})? This is an imagination of the global majority world from a vantage of an outsider. As Said highlighted, such images reduce places to a few symbols and stereotypes. Somalia is flattened to a set of images from boys on white sandy beaches, to streets-side vendors on dusty but neat roads. Sri Lanka prominently displays white sandy beaches and leisure, while other national associations such as the lotus flower, or its large population of elephants, are absent. Even Canada, situated within the global minority, is represented by a few stock images: moose, maple trees, cyclists and ice skating.

Scholars have argued that data sets are biased (Benjamin 2019; Buolamwini and Gebru 2018) based on knowledge systems with lineage in colonialism. Looking to the images, we propose a system trained on data that is Orientalist in its orientation and representation of the world, will, despite algorithmic adjustments and upgrades, at its core, perpetuate past injustices. The consistency of outputs indicates a persistent foundational association between a place and a representation. This forms the first part of our argument: the images, across countries and over time, suggest that a system fed on datasets that are Orientalist, produce and uphold the worldview that has created it.

\begin{figure}[htbp]
\centering
\includegraphics[width=0.55\textwidth]{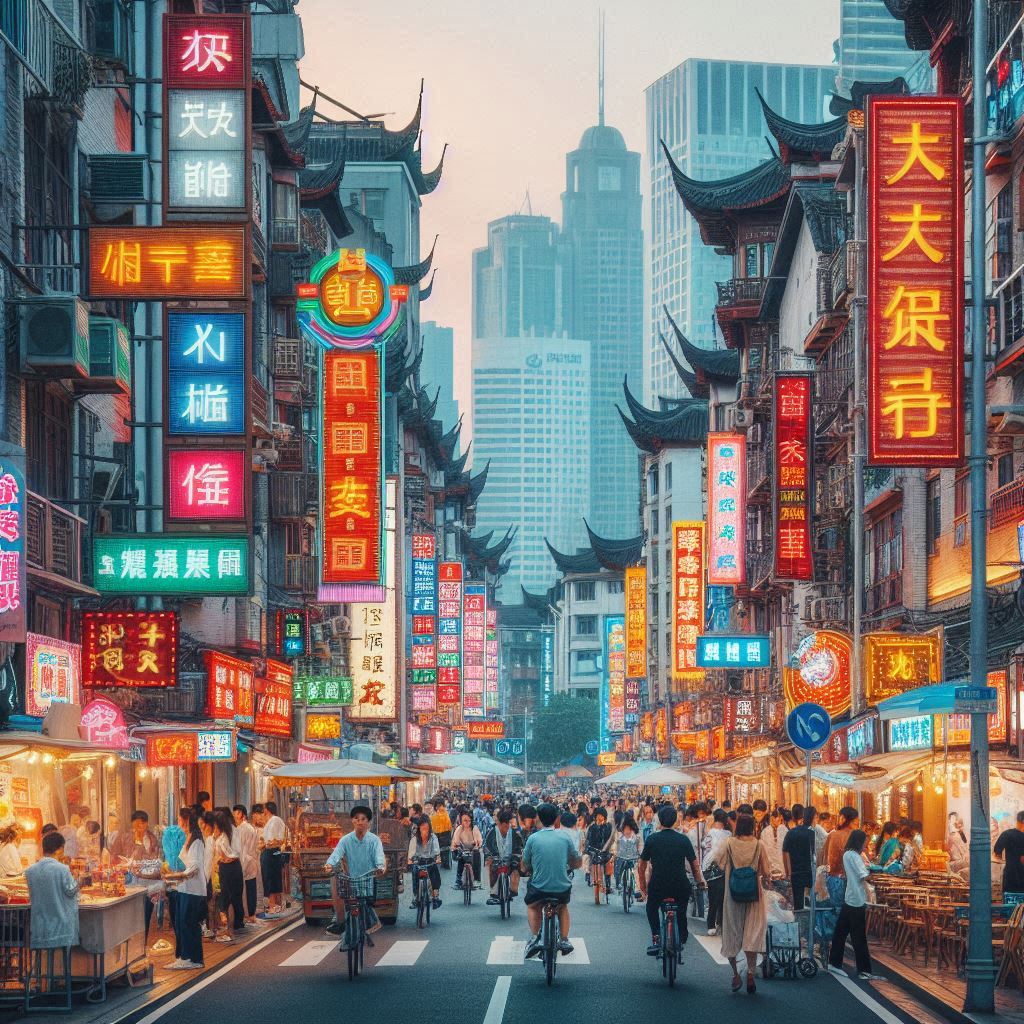}
\caption{``\nl{\includegraphics[height=\nlheight]{scripts/zh1.pdf}}'' {[}China{]} (21 July 2024)}
\label{fig:6}
\end{figure}

Consider the representations of China, which are presented through a limited repertoire of signifiers: lanterns, tourist landmarks, vertical neon signs with writing resembling (but actually nonsensical) Chinese characters, pandas, pagodas, dragons and bikes (Figure~\ref{fig:6}). Twenty-four out of twenty-eight images for the prompt ``China'' or ``\nl{\includegraphics[height=\nlheight]{scripts/zh1.pdf}}'' depict a bustling street scene in warm colours, combining pedestrians and riders on bikes or motorbikes between rows of buildings. In seventeen of the images for these prompts, the main feature are red lanterns, often occupying the top half of the image; in seven images, these are replaced by vertical street signs in neon colours, featuring nonsensical characters.

For the prompt ``Japan'', the dominant image is a temple garden with a pond and a red bridge. Stereotypical elements render the images recognisable as Japan, most commonly pink cherry blossoms. In eleven out of twenty-six images for both linguistic versions of ``Japan'', the pond contains koi carp (nishikigoi), traditional symbols of strength and longevity. Fifteen of the twenty images for ``Japan'' do not feature people at all.

\begin{quote}
Orientalism is not about objective knowledge of the East. Instead, it's a system of representation that creates a particular image of the East, often based on stereotypes and preconceived notions.
\end{quote}

These are not our words. They are taken verbatim from an AI-generated Google search (23 July 2025, 10:45 BST-British Summer Time). If we take this at face value and propose the images produced by Bing Create fall into an Orientalist view of the world, then the images of China and Japan rely on stereotypes that are easily distinguishable: cherry blossoms for Japan, red lanterns for China.

AI optimists could respond with the proposition that with more data and further training, outputs will become more precise in their statistical predictions. At the level of undefined prompts, our methodology suggests otherwise. As we repeatedly keyed in the same prompt structures, we were confronted with a system lacking variety and variation. Faced with an undefined prompt, the model consistently reverted to a small set of images. If prompted for an image of China, outputs converged on red lanterns or neon signs, for Japan, pink cherry blossoms.

\subsection{Fiction intermingled with fact}

As the images incorporated easily identifiable stereotypes, on closer scrutiny, aspects of these depictions emerged as false or misrepresentative. For Japan, temples feature in the cityscapes, but the architecture does not match the location; for example, temple forms that would be built in secluded places or on mountain tops appear in the middle of cities. Also incorrect, or nonsensical, are Chinese, Korean and Japanese scripts that appear in several images. The model is not designed to create a factual representation of a Chinese character or Japanese architecture, but to approximate the most likely visual compilation for a prompt.

\begin{figure}[htbp]
\centering
\includegraphics[width=0.55\textwidth]{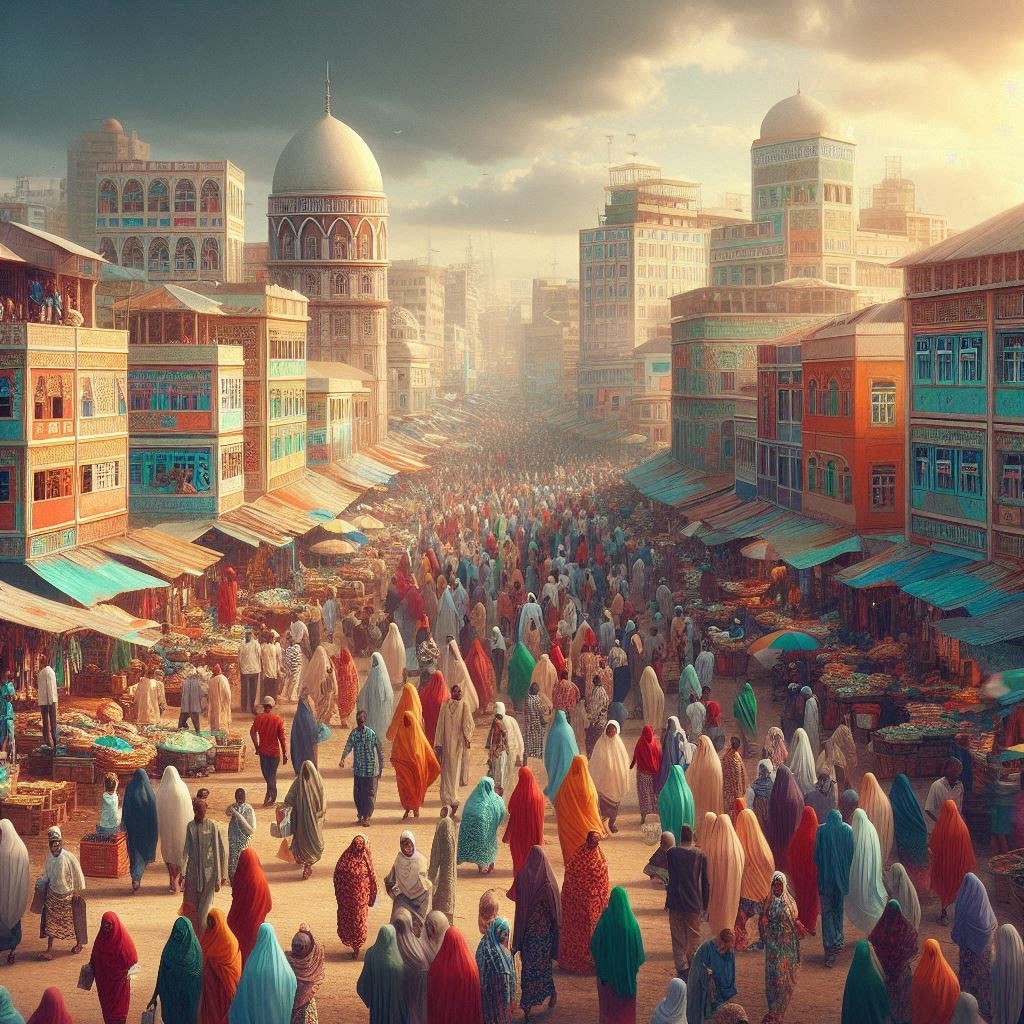}
\caption{``Soomaaliya + Magaalo'' {[}Somalia + city{]} (9 September 2024)}
\label{fig:7}
\end{figure}

This is repeated across countries and input languages. For Somalia, with its longstanding political conflicts, outputs lean towards an idyllic view, with pictures that are somewhat reminiscent of the palm tree lined coastline of a small (southern) part of Somalia's lengthy coastline stretch, but more so the coastal tourist spots of neighbouring Kenya. When English and Somali prompts for city/magaalo were added, Bing Create generated pictures of crowded street scenes of markets and Islamic/Arabic architecture more reminiscent of Yemen than the Horn of Africa. People dominated seemingly pedestrianised streets that tended to stretch to a central vanishing point, and motor vehicles were much less prevalent than in actuality in urban Somalia (Figure~\ref{fig:7}). This is not a representation of a Somali city, it is a representation of an imagination of a Somali city.

\begin{figure}[htbp]
\centering
\includegraphics[width=0.55\textwidth]{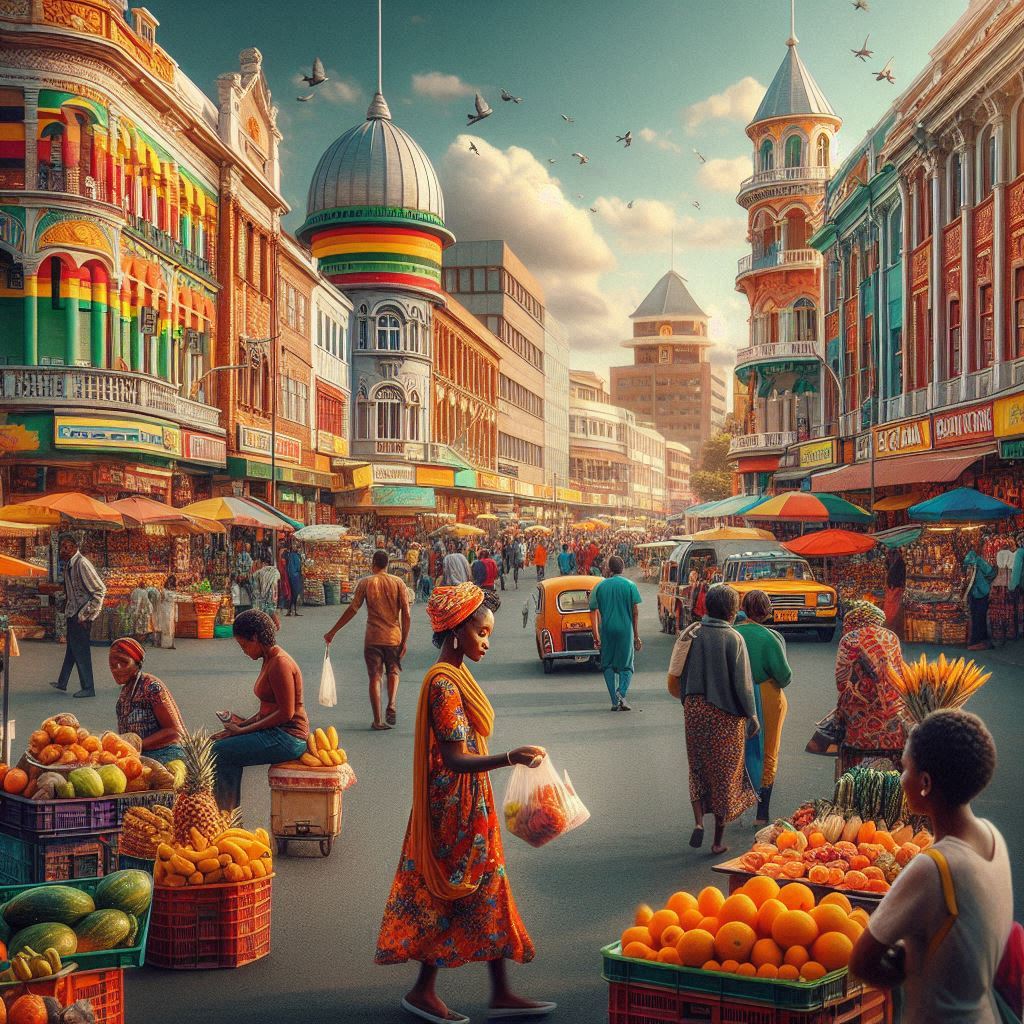}
\caption{``Zimbabwe + city'' (18 July 2024)}
\label{fig:8}
\end{figure}

\begin{figure}[htbp]
\centering
\includegraphics[width=0.55\textwidth]{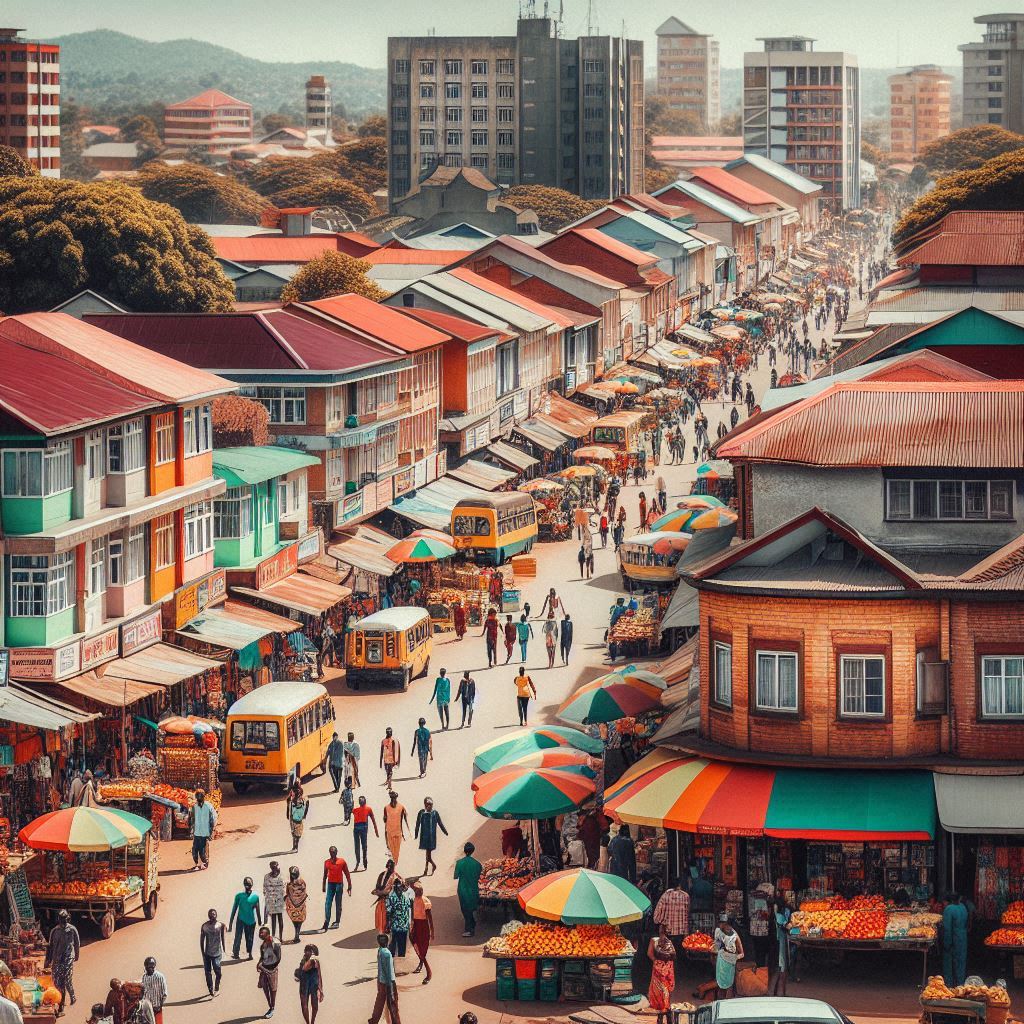}
\caption{``Zimbabwe + city'' (8 August 2024)}
\label{fig:9}
\end{figure}

In Zimbabwe, when prompted to generate images on ``Zimbabwe'' + ``City'', Bing's image generator predominantly produced street scenes blending colonial-era architecture with generic urban imagery (Figure~\ref{fig:8}). It incorporated visual elements more typical of South or East Asian, and Indian cities. In the earlier weeks, outputs showed some cultural accuracy, such as recognisable downtown Harare skylines and somewhat realistic vendor and street stalls. However, overall, most images misrepresented everyday life. Common inaccuracies included yellow school buses and vintage taxis (Figure~\ref{fig:9}) -- both not found in Zimbabwe -- or dressing people in Indian saris or West African agbadas. Market scenes consistently showed street vendors selling only fruits and vegetables, while on Zimbabwean city streets vendors sell clothing, electronics, and other goods. The streets, architecture, and people lacked specificity, depicting an assemblage of global `developing cities'.

\subsection{A tourist's fantasy}

The tool generated images that function as a consumable product for global audiences. Amidst inaccuracies, stereotypes and symbols, outputs most often depicted such varied localities as South Africa, Somalia, Zimbabwe and Sri Lanka as destination of sun, sea, sand and football. The insidiousness of Orientalism does not emerge in images that confine the Global Majority to poverty or conflict. Rather, it is in images that would not be out of place in a commercial tourist brochure, a voyeuristic representation that is reflective of western desires and global capitalist mediascapes (Urry and Larsen, 2011: 195-6).

\begin{figure}[htbp]
\centering
\includegraphics[width=0.55\textwidth]{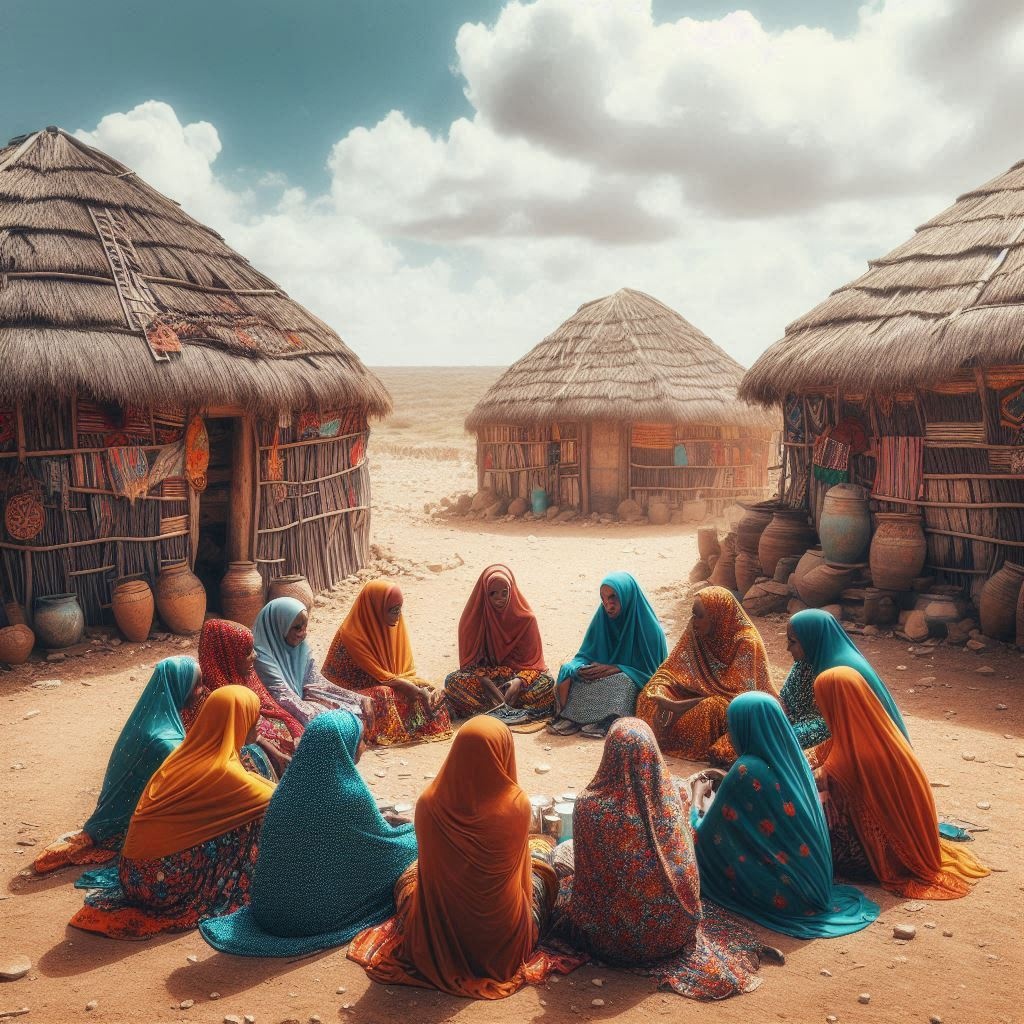}
\caption{``Somalia'' (27 July 2024)}
\label{fig:10}
\end{figure}

\begin{figure}[htbp]
\centering
\includegraphics[width=0.55\textwidth]{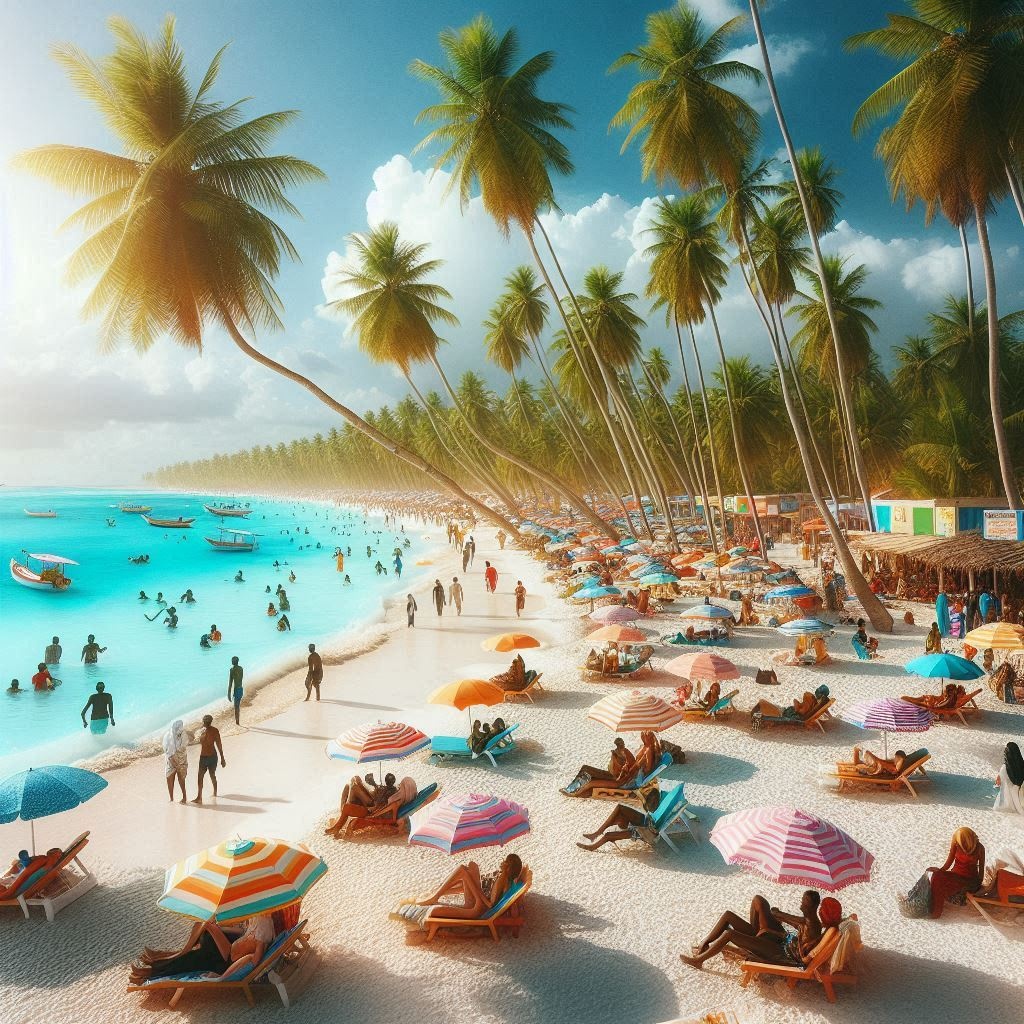}
\caption{``Soomaaliya'' {[}Somalia{]} (19 August 2024)}
\label{fig:11}
\end{figure}

For Somalia, the single word (in English) country prompt almost exclusively led to representations of tropical beaches filled with (mostly) young boys playing football. With palm trees, rustic beach huts and fishing boats (along with occasional camel), images gave an idyllic -- and generic -- view of a country that is often visually portrayed in global media in relation to conflict and humanitarian crisis. Intriguingly, it was only the English `Somalia' prompt (in comparison to `Soomaaliya') that generated an outlier in the Somalia dataset (Figure~\ref{fig:10}). This image showed a dusty village scene, with women seated in a circle. The image featured somewhat realistic depictions of wood and grass-roofed huts, as well as the women's dress. The scene is arid and is evocative of a dry period of scarcity. Beyond this single image, the outputs reverted to colourful beach scenes (e.g.\ Figure~\ref{fig:11}). 

\begin{figure}[htbp]
\centering
\includegraphics[width=0.55\textwidth]{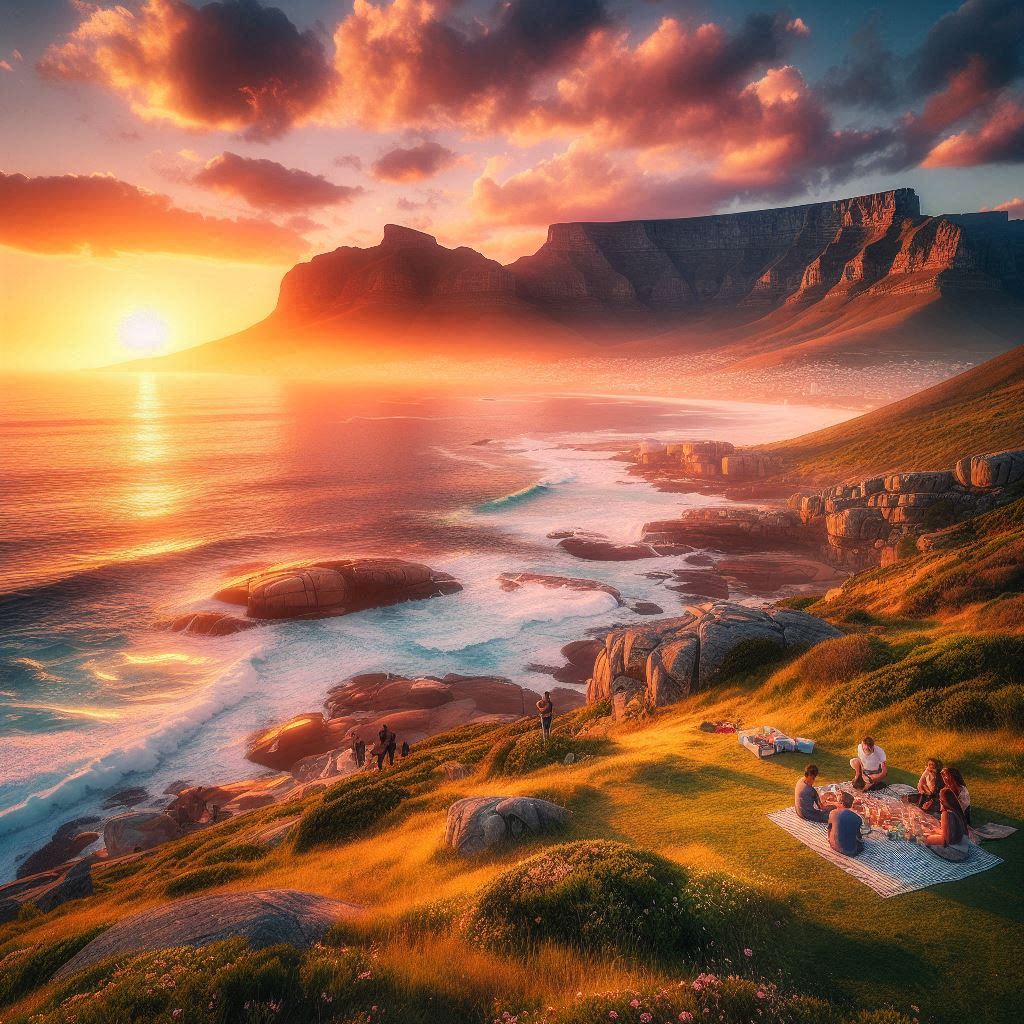}
\caption{``Cape Town'' (7 October 2024)}
\label{fig:12}
\end{figure}

\begin{figure}[htbp]
\centering
\includegraphics[width=0.55\textwidth]{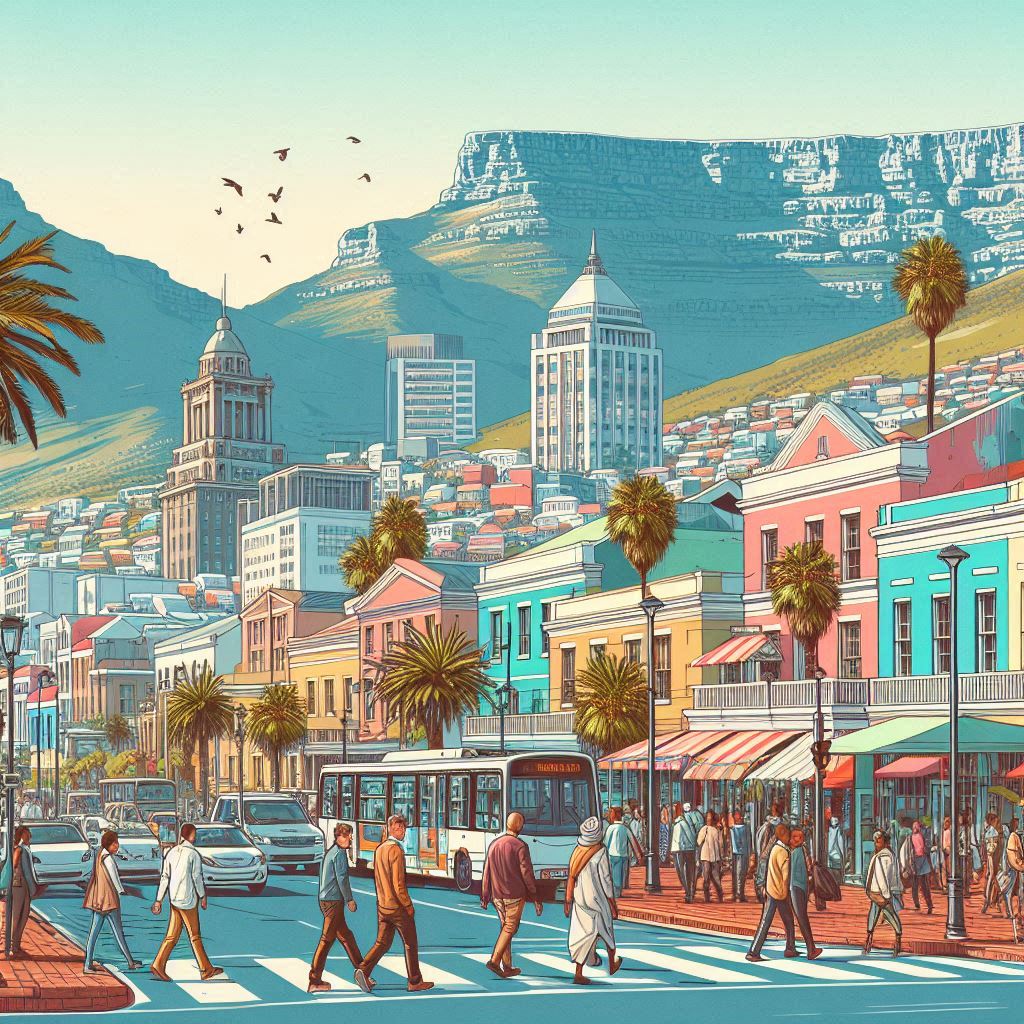}
\caption{``Stad Kaapstad'' {[}Cape Town{]} (4 November 2024)}
\label{fig:13}
\end{figure}

South Africa, too, was consistently represented as an idyllic destination for the western tourist, with wildlife, beaches and football. When broken down by its two iconic cities, Johannesburg and Cape Town, this superimposition takes on distinct tones, retaining an idealised vision, but reflecting the cities' layered histories within Orientalist imaginaries. When prompted in English, images of Cape Town appear to directly address the foreign tourist, depicting cable carts above the ocean, or young people sitting on grass, with Table Mountain and waves in the distance (Figure~\ref{fig:12}). The city disappears or fades in the background. When prompted in Afrikaans, Bing returns images of city, or more precisely, town life, transformed into an almost Europeanised streetscape, with rows of brightly painted heritage houses. The city is rendered as orderly, affluent, and intimate, closer to a Mediterranean resort town than a major African metropolis (Figure~\ref{fig:13}). The imagery emanating from Cape Town seems to spill over to less touristy Johannesburg. This is Johannesburg filtered through a Cape Town aesthetic.

\begin{figure}[htbp]
\centering
\includegraphics[width=0.55\textwidth]{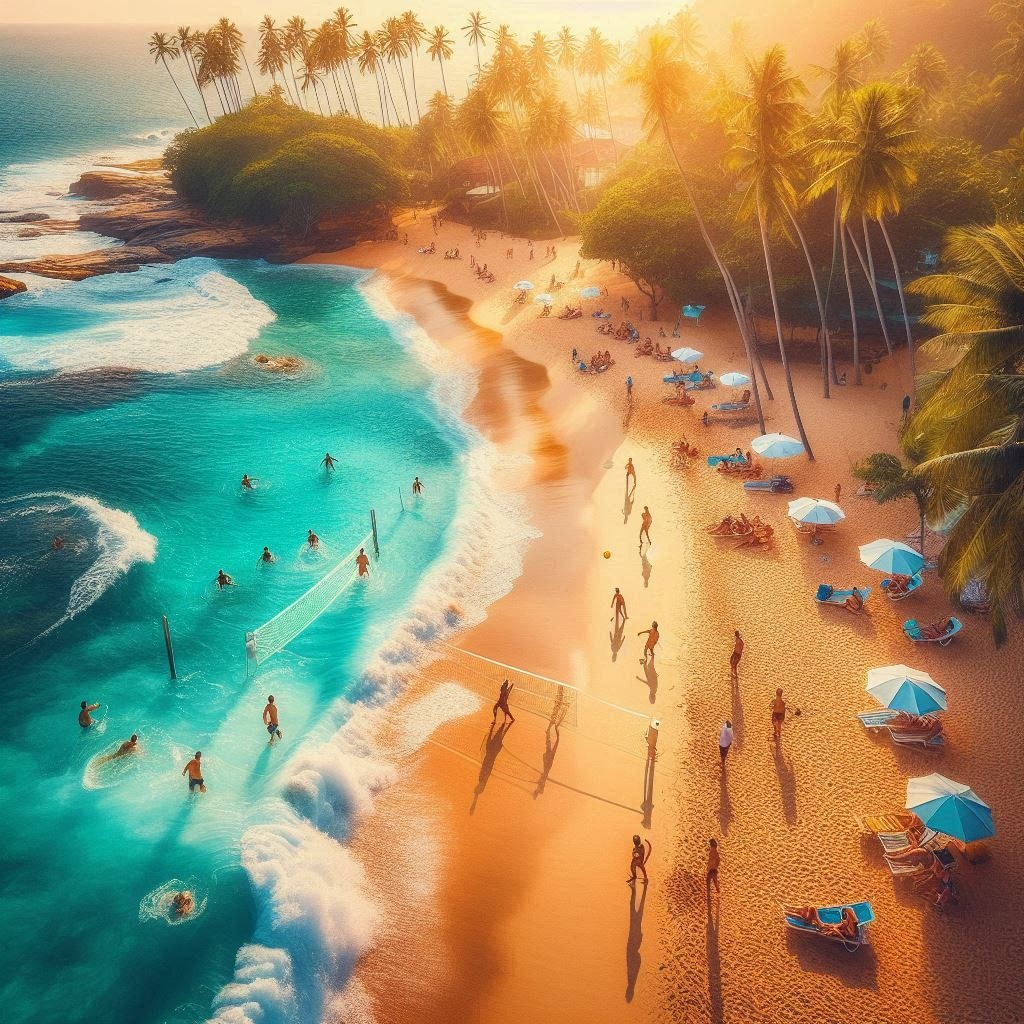}
\caption{``Ilaṅkai'' {[}Sri Lanka{]} (5 August 2024)}
\label{fig:14}
\end{figure}

With these projections of a western gaze generated as representative of African countries, the images depicting Sri Lanka give further pause. The prompt ``Sri Lanka'' similarly delivered images depicting a tropical paradise. Seemingly generated from the vantage point of a voyeur on a raised mountain side (or drone), the images display sandy beaches and an emerald sea. The tourist gaze is completed by scenes of sea-bathing, volleyball, cricket and football, played by mostly white people (Figure~\ref{fig:14}). Sri Lanka is a popular destination for western tourists, and, in the probability quotient of AI image generation, such an image is probable. Equally, Sri Lanka has suffered tremendous upheaval post-independence, including a twenty-six-year civil war, violent uprisings and terrorism (Amarasingam, 2019; Weiss, 2011). Politics, violence and strife, it appears, have been airbrushed out. It is possible - and likely - that this effect is informed by the model's guardrails. However, analytically, this is not only omission but a condition of possibility. The developmental and touristic imaginary depends on the bracketing of conflict, labour, extraction and governance. What the system cannot easily visualise - for example unequal property relations, coercive labour, contested sovereignty - is precisely what would trouble the coherence of the image as a consumable `place'.

\begin{figure}[htbp]
\centering
\includegraphics[width=0.55\textwidth]{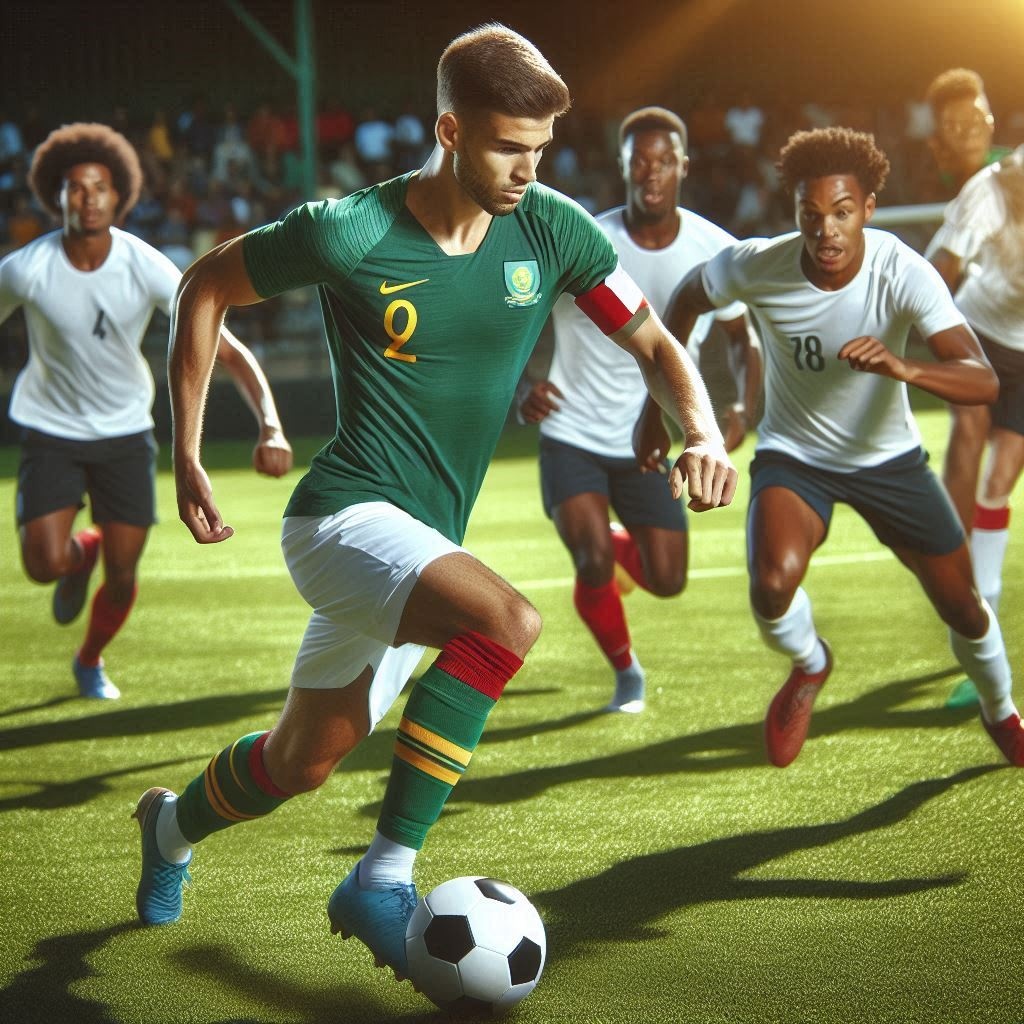}
\caption{``Iningizimu Afrika + idolobha'' {[}South Africa + city{]} (21 August 2024)}
\label{fig:15}
\end{figure}

Covert racism is embedded in several images. The depiction of South Africa, like Sri Lanka and Somalia, has wild animals, beaches and boys playing football. Some are of children chasing after a ball. Others are of grown-ups playing football, dressed in green jerseys, playing, most probably, professionally. In the images of the boys, they are black, with a white senior boy or adult in the background in a few instances, looking out for them, benign and smiling. In a case of the older boys or men playing what seems to be professional football, the ball is with the white person with black athletes chasing, trying to catch up (Figure~\ref{fig:15}).

\subsection{One generic future}

\begin{figure}[htbp]
\centering
\includegraphics[width=0.9\textwidth]{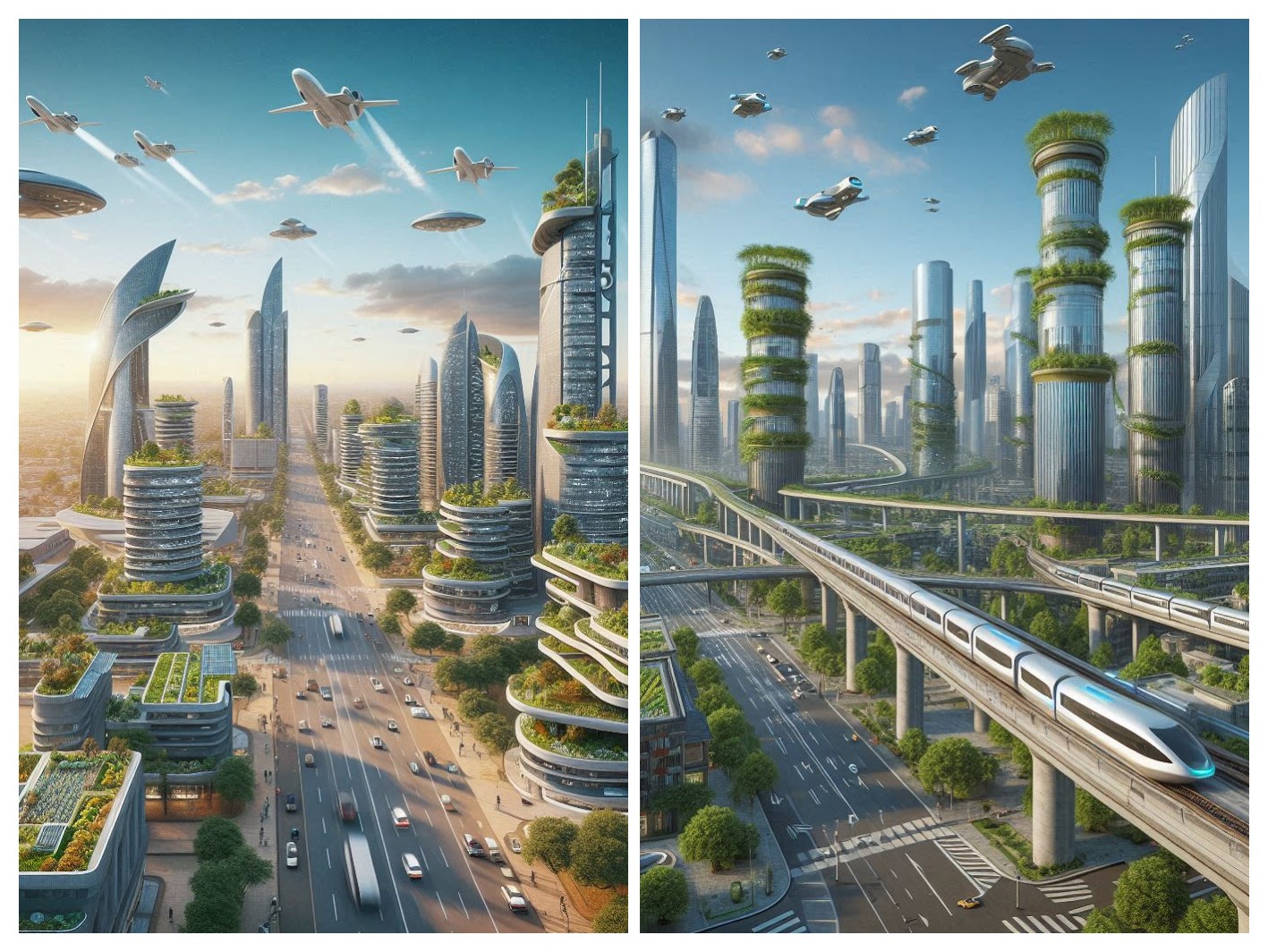}
\caption{Left: ``Zimbabwe + city + future'' (18 July 2024); Right: ``China + city + future'' (28 July 2024)}
\label{fig:16}
\end{figure}

\begin{figure}[htbp]
\centering
\includegraphics[width=0.9\textwidth]{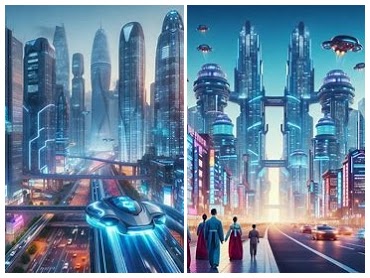}
\caption{Left: ``South Korea + city + future''; Right: ``North Korea + city + future'' (18 August 2024)}
\label{fig:17}
\end{figure}

When adding in the prompt ``Future,'' the `catching up' of the majority world seems largely to have happened. The future imagined by Bing Create is strikingly similar across countries and continents. In slightly under 90 per cent of the images (298/333), while still tinged with the same country signifiers, the prompt ``future'' resulted in shiny vehicles, urban gardens and sleek high-rises (e.g.\ Figure~\ref{fig:16}). In east and southeast Asia, images' colour palate sometimes shifts from cyan/aqua to warmer red/purple hues, but the tall sleek buildings, clean streets and vehicles remain (Figure~\ref{fig:17}). The colour-scheme of Canada's present-day city was the closest to that of the generic future.

\begin{figure}[htbp]
\centering
\includegraphics[width=0.55\textwidth]{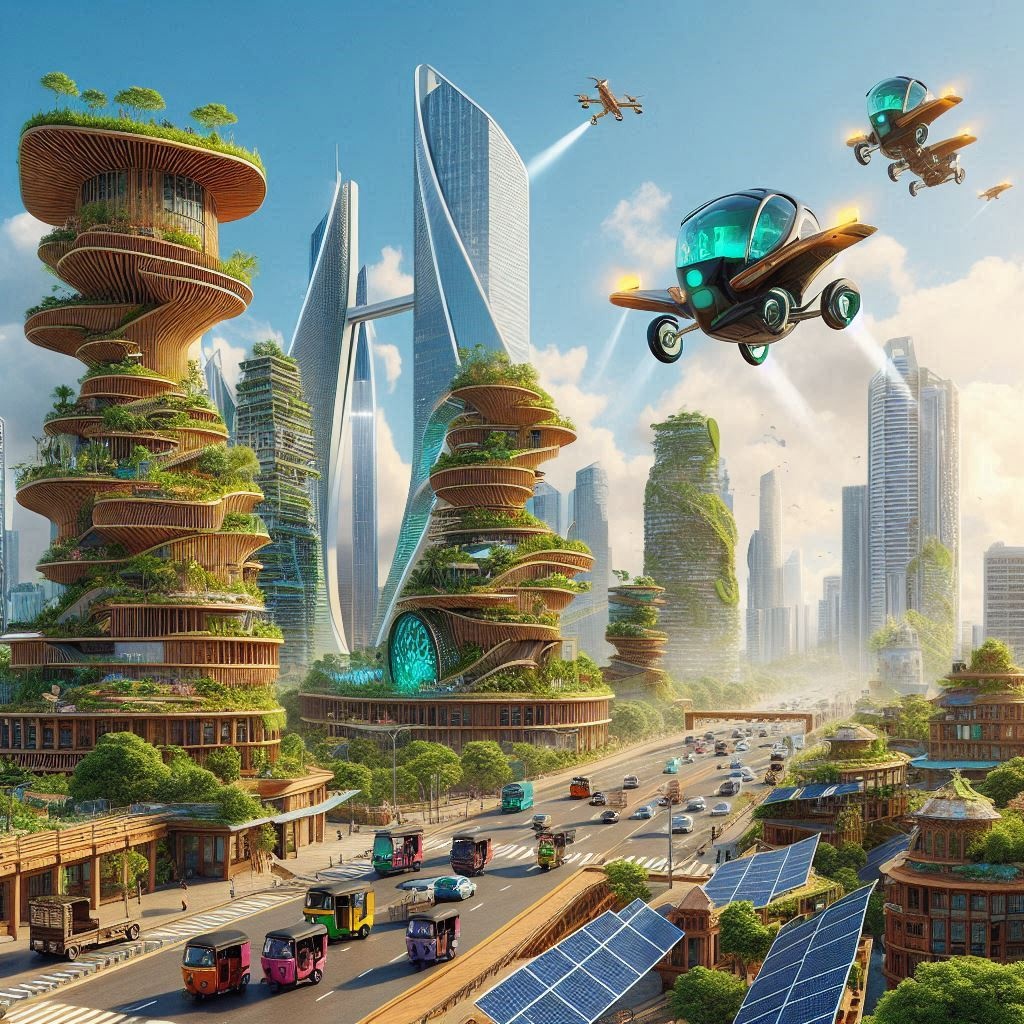}
\caption{``śrī laṁkā + nagarayē + anāgataya'' (12 July 2024)}
\label{fig:18}
\end{figure}

Be it India, Somalia or Canada, what appears is not simply `a' future, but a modernisation script, that is, the glossy city as a universal endpoint, with difference retained as surface ornamentation. Take Sri Lanka: across the three linguistic variations, ``Sri Lanka'' + ``City'' + ``Future'' produces a homogenous vision that could be conceptualised as `corporate steampunk'. Embodied by airborne tuk-tuks---South Asian three-wheeler taxis but with jet propulsion---that seem to pepper the sky of the Sri Lanka City Future, this imagery appears to infuse futuristic transport possibilities with twentieth century technologies (Figure~\ref{fig:18}). An additional curiosity includes how the tool reorientates gender. Women are represented wearing traditional sarees, while men are in contemporary western suits. Is the model predicting a future that accelerates western sartorial colonialism over the male body but not the woman's body, or is it reflecting deep-set patterns of misogyny?

In images from southern Africa, the singular future aesthetic is integrated with stereotypical imagery and fiction of the African cities. References to Harare's tree-lined streets or the iconic presence of Table Mountain in Cape Town are visible amidst skyscrapers, multilevel urban gardens, solar panels and flying vehicles. In some images, the stereotypical football field is present, but modern instead of dusty patches previously depicted. Other cultural or architectural markers are absent. In Zimbabwe, in some cases, features more reminiscent of east/southeast Asia (Malay, Thai, Chinese) dominated, including bamboo buildings, sari-wearing figures, and open-air markets dominated, retaining elements of fiction.

\begin{figure}[htbp]
\centering
\includegraphics[width=0.55\textwidth]{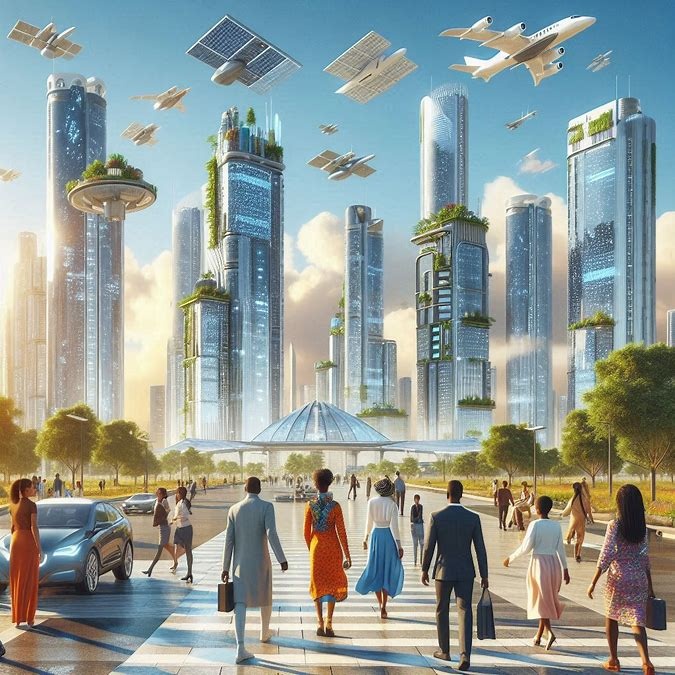}
\caption{``Ethiopia + city + future'' (30 July 2024)}
\label{fig:19}
\end{figure}

\begin{figure}[htbp]
\centering
\includegraphics[width=0.55\textwidth]{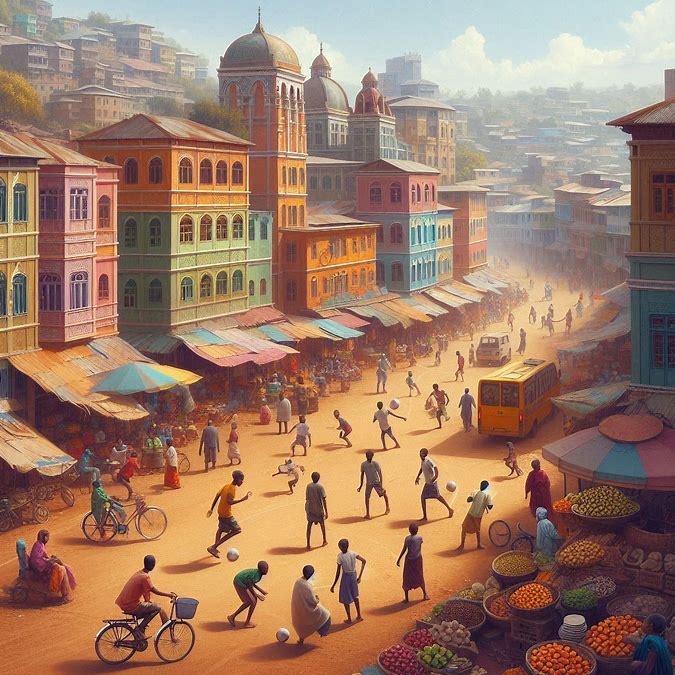}
\caption{``\nl{\includegraphics[height=\nlheight]{scripts/am3.pdf}}'' {[}Ethiopia + city +future{]} (30 July 2024)}
\label{fig:20}
\end{figure}

For a few countries, most consistently Ethiopia but also others in South Asia and Africa, the future city is emerging from the horizon, encroaching on the present-day city. Ethiopia's urban future was one of the distinct cases in which language affected results. In English, outputs followed the general pattern with a generic future aesthetic (Figure~\ref{fig:19}). In Amharic, the future hardly differed from the images of the present-day (Figure~\ref{fig:20}).

In sum, across the prompt structures, countries and languages, our experiment revealed a strong correlation of Orientalist data with Orientalist images: depicting an outsider's view, and reducing places to a few symbols, stereotypes and inaccuracies. The model's statistically predictive outputs leave little scope for varied realities of place, largely generating images for a (western) tourist. Such predictability of outputs was most stark when the future is prompted, generating a generic vision void of local futurisms.

\section{Discussion: Power, knowledge and the discourse of AI}

This section returns to our query into what this set of inputs and outputs reveal about the opaque operations of this generative AI tool, situating our images within Foucault's theory of discourse and Orientalism. We begin by considering, what do our images reveal about who gets to speak in the world of AI, and from which positions of power do they speak?

A partial answer is that it is the concentrated power of platform firms, investors, and technological elites who set agendas, standards, and narratives of AI capabilities (e.g. Guest 2025a, 2025b; Little and Winch, 2021). Alternatively, Said proposes representations must be understood in context. While generative AI is novel in scale and accessibility, it sits within longer histories of computation, automation and cybernetic imaginaries that shape how intelligence is defined and valued (Winterson, 2021). `AI' is a moment in a technological evolution that can be stretched back to the first Industrial Revolution. Beyond the hardware, AI's software is a product of several centuries of learning (Lawrence, 2024).

In \emph{the Birth of the Clinic}, Foucault (1973) explains how the discourse around medicine not only shaped medical practices but also how health and illness are understood and treated. In the context of the images, generative AI and its visionaries can be conceptualised as products of the discourse around science and technology. Scientific discourse carries with it the implicit understanding of human progress. Couched within a tech-optimist perspective, the allure of technological progress in the future fuels a belief that the follies of past and present practices will be mitigated. This belief proposes that generative AI will save time and save lives (Barasa et al., 2025), by synthesising existing information in a nano second to provide the most statistically plausible outcome.

This discourse of AI as the context for Bing Create brings us back to the prompts and images in our study. There are two things to note here. First, the images that Bing Create generated were based on learnings from the data that it was trained on, which, historically, has been dominated by western countries and populations. Second, those who developed the models that determined the probability quotient of these images are likely wrapped in the world of discourse around AI and technology. Such global concentration is not static: where data is produced is shifting, for example, with growing data production in China and India. Indeed, the political economy of data annotation has global dimensions (Wu et al., 2025). Yet we find this diversity is not reflected in the outputs. Our images indicate a persistent and pervasive influence of Orientalist/development discourse.

Looking at our images, the representations of the future have undoubtedly been informed by historical legacies of colonialism/capitalism, insofar as these have shaped the visual datasets that are drawn on. However, the potential significance of these generated images lies not simply in their discursive power, but also because it is generative AI itself that is being positioned as a vehicle of future development. The images generated of the future take the process of development to its idealised future: there are no more haves or have nots. All countries are included in one sanitized future, hardly distinguishable whether it is Sri Lanka, Indonesia or South Africa, albeit for flying tuk-tuks or traditional women's dress.

As this AI-generated imagery presents the fulfilment of the development promise, it also becomes caught in development's problematic underpinnings. These dominant discourses - of science and technology, development and AI - promise global progress, but their realities contradict this promise. Noted earlier, critical scholarship has highlighted the contradictions of development discourse. Escobar (1984: 378) explains how development, while purporting to be salvation for underdeveloped countries, maintains domination and economic exploitation, supported by discourse that prevents the articulation of alternatives. Sachs expands on this, suggesting the failure of development is rooted in a promise of an equal world without material differences. Sachs shows how equality through development is impossible, masking the ecological catastrophe of striving for such a future. Asking his readership to consider a world where every city is as rich, as illuminated, as Los Angeles, Sachs highlights the vast amount of waste such a scenario would produce. The discourse of development ignores the earth's limited resources, promising a future of expanding prosperity and equality that is illogical given earth's sustainability.

In Bing Create's images of the future, development's promise of equality reaches its apogee. Differences between the Orient and the Occident are minimised; countries are represented by variations of the same urban future. This future has no data centres, no hints of energy consumption, no electronic waste.

The discourse of development -- and the discourse of AI as revealed in our images -- is sustained by excluding and ignoring its material realities. If confronted with the reality of development in the world -- with growing inequality, extraction and waste -- its superficiality and impossibility are unambiguous. The generative AI model sustains the tenuous discourse of development in what is present and what is absent in the images: producing images of a singular, technologically-enabled future, even as the tool itself relies on inequalities and forms of extraction that make such a future impossible.

\section{Conclusion}

Our collective experiment set out to explore how -- working across diverse countries and languages -- we might gain insight into the black box of generative AI models. While focusing on one tool, we argue that the patterns that emerged across the images, over time, country and linguistic input, reveal an important reality about discourse and power that sit behind, and are produced through, the text-to-image results. First, the images consistently project imaginaries of mass tourism. When prompting the model to consider future cities, variation across countries disappears, producing a common aesthetic: bright, shiny and technological. Second, we argue that behind the monotony of the images is a perverse discourse that reflects the contradictions and impossibility of development. In the images, we see AI removing distinctions between haves and have nots, generating one future -- a future that hides the material realities of AI.

From here, we return to our initial query: What does this approach reveal about the nature and operations of colonial power in contemporary generative AI infrastructures? From one view, even as Bing Create reproduces and sits within the discourse of development -- with its divisions, inequalities and impossibilities -- this colonial present might be argued to have the potential to be benign. Generative AI is programmed to find patterns, not to necessarily be harmful or malicious. In fact, Bing Create has programmed guardrails to promote fairness. From another view - and this is what we have aimed to argue - it is precisely this banality, in the following of rules and patterns, that makes the generative AI tool capable of harm, generating images that erase plurality and obscure its own destructiveness.

In her reflections of the trial of Eichmann, Hannah Arendt put forward the argument that the potential for great evil sits in the banal:

\begin{quote}
He was not stupid. It was sheer thoughtlessness-something by no means identical with stupidity - that predisposed him to become one of the greatest criminals of that period. (Arendt, 1994: 287-8)
\end{quote}

We invoke Arendt not to collapse radically different orders of violence, but to name a shared mechanism: how patterned, procedural reproduction can enable harm without intent, precisely through the foreclosure of judgment and imagination. Thoughtlessness is not stupidity. The model is not stupid. Generative AI models can and will become more complex. They can encompass more data. They can become more precise. But the models are thoughtless. For Arendt, evil is possible precisely in the banal, when there is no thought, no openness to the possibility of unpredictability. Even if opaque in their operations (Burrell 2016), generative AI models are not unpredictable. When confronted with undefined prompts, the model predictably generated images that sustain and extend a discourse of Orientalism/development, obscuring its reliance on inequality and exploitation.

As academics, we sit within institutions whose physical and epistemological foundations reflect and reproduce colonial hierarchies of place, people and knowledge. Equally, our own personal and professional lives are becoming entwined with AI models with and without our intention or consent. We recognise the difficulty of sitting outside of the discourse of development and AI. Yet we also recognise the danger inherent in a single story, especially the story of development that obscures and accelerates the exploitation of people and planet.

Discourse is encompassing -- ever so as depicted in the singularity of how Bing Create generated images of the future - but it is not totalising (Escobar, 1984). Throughout the past centuries of colonialism and development, there remain diverse alternative forms of knowledge and ways of being, including that which is recognised and contended with in critical and decolonial scholarship in which this article sits. There are instances of experimental uses of generative AI that seek to unsettle the default banality of the machine, through, for example, efforts to critically curate datasets and carefully deployed and refined prompt engineering.\footnote{In one example, \href{https://vimeo.com/868332312/2f837b5f9b}{Noga Mo Jozi}, a collective of scholars and artists has built on a curated collection of artworks, rituals, architecture, lost or partially destroyed indigenous knowledge, and generative AI, to create a dreamscape of Johannesburg inspired by indigenous knowledge systems (South African Cities Network, 2024: 28). Their work points to potential avenues to contend with the `default' outputs of generative AI tools, a response to Wiredu's ``positive programme'' of decolonising by ``exploiting as much as is judicious the resources of our own indigenous conceptual schemes'' (Wiredu, 1996: 136).} Our study points to the importance of further research that grapples with the systems of power that surround generative AI, and what this implies for where and how alternative imaginaries are possible. Equally, it reinforces the view that discourses aimed at a more equitable future -- with or without AI -- cannot come from banally reproducing the past. Such a future must begin with efforts to expose and contend with the unequal systems of power that generative AI tools are both producing and being produced by.

\section*{Statements and declarations}

\paragraph{Acknowledgements.} We are grateful to Aarushi Gupta who participated in the data collection process and analysis workshops.

\paragraph{Declaration of conflicting interest.} The authors declare no potential conflicts of interest with respect to the research, authorship, and/or publication of this article.

\paragraph{Funding statement.} The workshops for this project were organised by the SOAS Centre for AI Futures, with King's College London's Department of Digital Humanities, and funded by SOAS, University of London.

\paragraph{Ethical approval.} Ethical approval was not required. The study did not involve data pertaining to human subjects.

\paragraph{Data availability statement.} The underlying data for this study is publicly available via \url{https://github.com/objetpetitm/SOAS_AI_images}.

\section*{References}
\addcontentsline{toc}{section}{References}
\begingroup
\small
\begin{list}{}{\setlength{\leftmargin}{1.5em}\setlength{\itemindent}{-1.5em}\setlength{\itemsep}{0.3em}\setlength{\parsep}{0pt}}
\item Adams R (2021) Can artificial intelligence be decolonized?. Interdisciplinary Science Reviews 46(1-2): 176-197.
\item Adams R, Alayande A, Brey Z et al. (2023) A new research agenda for African generative AI. \emph{Nature Human Behaviour} \emph{7}(11): 1839-1841.
\item Ademolu E (2023) Racialised representations of Black African poverty in INGO communications and implications for UK African diaspora: Reflections, lessons and recommendations. \emph{Journal of Philanthropy and Marketing} 28(4): e1718.
\item Ademolu E (2024) Laughing from the Outside-In: Considering `What's Up Africa' as an(other) humorous humanitarian digilantism. \emph{Journal of International Development} 36(5): 2393--2413.
\item Ademolu E (2025) Don't Wait, ACT! A Culturally Competent Approach for Inclusive Development Communications with African Diaspora Communities. \emph{Journal of Philanthropy} 30(3): e70024.
\item Alenichev A, Kingori P and Grietens KP (2023) Reflections before the storm: the AI reproduction of biased imagery in global health visuals. \emph{The Lancet Global Health} 11(10): e1496-e1498.
\item Ali, Syed Mustafa (2024) AI and/as Racialised Political Theology. \emph{Philosophy and Society} 35(3): 645--666.
\item Amarasingam A (2019) Terrorism on the Teardrop Island: Understanding the Easter 2019 Attacks in Sri Lanka. \emph{CTC Sentinel} 12(5): 1-10.
\item Anwar MA and Graham M (2020) Hidden transcripts of the gig economy: labour agency and the new art of resistance among African gig workers. \emph{Environment and Planning A: Economy and Space} 52(7): 1269-1291.
\item Arendt H (1994) \emph{Eichmann in Jerusalem: A Report on the Banality of Evil} (revised and enlarged edition). London: Penguin Books.
\item Arjana SR (2025) \emph{Colonial Geographies, Tourist Imaginaries, and Mystical Landscapes} (1\textsuperscript{st} edition). New York: Routledge.
\item Barasa H, Scales J, Large O and Iosad A (2025) \emph{Governing in the Age of AI: Unlocking a New Era of Transformation in Africa}. Report, Tony Blair Institute for Global Change, 3 April. Available at: \url{https://institute.global/insights/politics-and-governance/governing-in-the-age-of-ai-unlocking-a-new-era-of-transformation-in-africa} (accessed 27 January 2026).
\item Benjamin R (ed) (2019) \emph{Captivating technology: Race, carceral technoscience, and liberatory imagination in everyday life}. Durham and London: Duke University Press.
\item Buolamwini J and Gebru T (2018) Gender shades: Intersectional accuracy disparities in commercial gender classification. \emph{Proceedings of Machine Learning Research} 81: 77-91.
\item Campbell D (2012) The Iconography of Famine. In: Batchen G, Gidley M, Miller NK and Prosser J (eds) \emph{Picturing Atrocity: Photography in Crisis.} London: Reaktion Books, pp. 79--91.
\item Carabantes M (2020) Black-box artificial intelligence: an epistemological and critical analysis. \emph{AI and Society} 35(2): 309-317.
\item Cave S (2020) The problem with intelligence: its value-laden history and the future of AI. In \emph{Proceedings of the AAAI/ACM Conference on AI, Ethics, and Society,} February 2020, pp. 29-35.
\item Chatterjee T (2018) Colonial Discourse Analysis: Foucault's Power/Knowledge Nexus and Said's Orientalism. \emph{Journal of Humanities \& Social Science} 7(1): 199-208.
\item Chonka P (2019) The empire tweets back?\# HumanitarianStarWars and memetic self-critique in the aid industry. \emph{Social Media+ Society} 5(4).
\item Chonka P, Diepeveen S and Haile Y (2023) Algorithmic power and African indigenous languages: search engine autocomplete and the global multilingual Internet. \emph{Media, Culture \& Society} 45(2): 246-265.
\item Chouliaraki L (2006) \emph{The Spectatorship of Suffering}. London: Sage.
\item Chouliaraki L (2010) Post-Humanitarianism: Humanitarian Communication Beyond a Politics of Pity. \emph{International Journal of Cultural Studies} 13(2): 107--126.
\item Colombo G, Niederer S, de Gaetano C and M Borie (2023) Prompting Generative Visual AI for Biodiversity: From Prompt Engineering to Prompt Design. In: \emph{Generative Methods -- AI as Collaborator and Companion in the Social Sciences and Humanities Conference}, Aalborg University, December 6--8.
\item Couldry N and Mejias UA (2019) Data colonialism: Rethinking big data's relation to the contemporary subject. \emph{Television \& New Media} 20(4): 336-349.
\item Crawford K (2021) \emph{The atlas of AI: Power, politics, and the planetary costs of artificial intelligence}. New Haven: Yale University Press.
\item de Seta G (2024) Synthetic Probes: A Qualitative Experiment in Latent Space Exploration. \emph{Sociologica} 18(2): 9--23. doi: 10.6092/issn.1971-8853/19512.
\item de Seta G, Pohjonen M and Knuutila A (2024) Synthetic ethnography: Field devices for the qualitative study of generative models. \emph{Big Data \& Society} 11(4): p.20539517241303126.
\item Dogra N (2012) \emph{Representations of global poverty: Aid, development and international NGOs}. London: I. B. Tauris.
\item El Aidi A and Yechouti Y (2017) Michel Foucault's theory of discourse in Edward Said's theory of Orientalism. \emph{The Criterion: An International Journal in English} 8(4): 1059-1072.
\item Elish MC and boyd d (2018) Situating methods in the magic of Big Data and AI. \emph{Communication monographs} 85(1): 57-80.
\item Escobar A (1984) Discourse and power in development: Michel Foucault and the relevance of his work to the Third World. \emph{Alternatives} 10(3): 377-400.
\item Escobar A (1995) \emph{Encountering development: The making and unmaking of the Third World.} Princeton: Princeton University Press.
\item Ferguson J (1994) \emph{The Anti-Politics Machine: Development, Depoliticization, and Bureaucratic Power in Lesotho}. Minneapolis: University of Minnesota Press.
\item Foucault M (1973) \emph{Birth of the Clinic} (trans. AM Sheridan-Smith). London: Tavistock.
\item Foulds J, Feldman P and Pan S (2025) Grok's antisemitic rant shows how generative AI can be weaponized. \emph{The Conversation}, 18 June. Available at: \url{https://doi.org/10.64628/AAI.59rg57mky} (accessed 27 January 2026).
\item Gorska AM and Jemielniak D (2023) The invisible women: uncovering gender bias in AI-generated images of professionals. \emph{Feminist Media Studies} 23(8): 4370--4375.
\item Gray C (2023) More than extraction: Rethinking data's colonial political economy. \emph{International Political Sociology} 17(2): p.olad007.
\item Guest O and van Rooij I (2025). Critical Artificial Intelligence Literacy for Psychologists. \emph{PsyArXiv}. Preprint 4 October 2025. https://doi.org/10.31234/osf.io/dkrgj\_v1.
\item Guest O, Suarez M, van Rooij I (2025) Towards Critical Artificial Intelligence Literacies. \emph{Zenodo.} Preprint 2 December 2025. \url{https://doi.org/10.5281/zenodo.17786243}.
\item Hall S (1992) The West and the Rest: Discourse and Power. In: D Morley (ed) \emph{Essential Essays (Volume 2).} Durham, NC: Duke University Press, pp. 185-224.
\item Keenaghan N and Reilly K (2017) Mediating representations of poverty and development: a help or a hindrance? \emph{Third World Quarterly} 38(6): 1272--1290.
\item Kerche FW, Zook M and Graham M (2026) The silicon gaze: A typology of biases and inequality in LLMs through the lens of place. \emph{Platforms \& Society} 3.
\item Kerr, D (2025) Musk's AI Grok bot rants about `white genocide' in South Africa in unrelated chats. \emph{The Guardian}, 15 May. Available at: \url{https://www.theguardian.com/technology/2025/may/14/elon-musk-grok-white-genocide} (Accessed: 4 February 2026).
\item Kothari U (2015) Reworking colonial imaginaries in post-colonial tourist enclaves. \emph{Tourist Studies} 15(3): 248-266.
\item Kotliar DM (2020) Data orientalism: on the algorithmic construction of the non-Western other. \emph{Theory and Society} 49(5/6): 919--939.
\item Lawrence ND (2024) \emph{The atomic human: Understanding ourselves in the age of AI}. London: Random House.
\item Lewis D, Rodgers D and Woolcock M (eds) (2014) \emph{Popular Representations of Development: Insights from Novels, Films, Television and Social Media} (1st edition). Abingdon: Routledge.
\item Lewis JE, Whaanga H and Yolgörmez C (2024) Abundant intelligences: placing AI within Indigenous knowledge frameworks. \emph{AI and Society} 40: 2141--2157.
\item Lidchi H (1999) Finding the right image: British development NGOs and the regulation of imagery. In: T Skelton and T Allen (eds)\emph{ Culture and Global Change}. Abingdon: Routledge.
\item Little B and Winch A (2021) \emph{The new patriarchs of digital capitalism: Celebrity tech founders and networks of power}. Abingdon: Routledge.
\item Locke LG, Hodgdon G (2024) Gender bias in visual generative artificial intelligence systems and the socialization of AI. \emph{AI and Society} 40: 2229--2236.
\item Mamdani M (1996) \emph{Citizen and Subject: Contemporary Africa and the legacy of late colonialism}. Princeton: Princeton University Press.
\item Mohamed S, Png MT and Isaac W (2020) Decolonial AI: Decolonial Theory as Sociotechnical Foresight in Artificial Intelligence. \emph{Philosophy and Technology} 33: 659--684.
\item Muldoon J and Wu BA (2023) Artificial intelligence in the colonial matrix of power. \emph{Philosophy and Technology} 36(80): 1-24.
\item Natale S (2021) \emph{Deceitful media: Artificial intelligence and social life after the Turing test.} Oxford: Oxford University Press.
\item Plantin JC, Lagoze C, Edwards PN and Sandvig C (2018) Infrastructure studies meet platform studies in the age of Google and Facebook. \emph{New media \& society} \emph{20}(1): 293-310.
\item Racevskis K (2005) Edward Said and Michel Foucault: Affinities and Dissonances. \emph{Research in African Literatures} 36(3): 83-97.
\item Richey LA (2016) Tinder humanitarians: The moral panic around representations of old relationships in new media. \emph{Javnost---The Public} 23(4): 341--398.
\item Rostow WW (1960) \emph{The Stages of Economic Growth: A Non-Communist Manifesto.} Cambridge: Cambridge University Press.
\item Sachs W (ed) (2019) \emph{The development dictionary: a guide to knowledge as power} (3\textsuperscript{rd} edition). London: Zed Books.
\item Said EW (1978) \emph{Orientalism}, New York: Pantheon Books.
\item Schwarz KC and Richey LA (2019). Humanitarian humor, digilantism, and the dilemmas of representing volunteer tourism on social media. \emph{New Media \& Society} 21(9): 1928--1946.
\item South African Cities Network (2024) \emph{The Working City: Actions for Growth and Recovery}. Report for the 2023 Urban Festival. South Africa: South African Cities Network.
\item Toyama, K. (2015). \emph{Geek heresy: rescuing social change from the cult of technology}. New York: PublicAffairs.
\item Urry J and Larsen J (2011) \emph{The Tourist Gaze 3.0}. London: Sage Publications.
\item Van Dijck J (2013) `You have one identity': Performing the self on Facebook and LinkedIn. \emph{Media, Culture \& Society} 35(2): 199-215.
\item Weiss G (2011) \emph{The Cage: The Fight for Sri Lanka and the Last Days of the Tamil Tigers.} London: Vintage Books.
\item Winterson J (2021) \emph{12 Bytes: How artificial intelligence will change the way we live and love}. London: Random House.
\item Wiredu K (1996) \emph{Cultural Universals and Particulars}. Indianapolis: Indiana University.
\item Wong M (2025) Elon Musk Updated Grok. Guess What It Said? \emph{The Atlantic}, 11 July. Available at: https://www.theatlantic.com/technology/archive/2025/07/new-grok-racism-elon-musk/683515/ (Accessed: 3 January 2026).
\item Wu T, Muldoon J and Xia B (2025) Global data empires: Analysing artificial intelligence data annotation in China and the USA. \emph{Big Data \& Society} 12(2): 1-13.
\end{list}
\endgroup

\end{document}